# Novel Inertial Self-Assembly Dynamics and Long-Range Spatial Ordering of Interacting Droplet Ensembles in Confined Microfluidic Flows

Wenyang Jing[1] and Hee-Sun Han[1,2,3*]

1. Center for Biophysics and Quantitative Biology, University of Illinois at Urbana-Champaign, Urbana, IL, 61801
2. Department of Chemistry, University of Illinois at Urbana-Champaign, Urbana, IL, 61801
3. Carl R. Woese Institute for Genomic Biology, University of Illinois at Urbana-Champaign, 1206 W. Gregory Dr, Urbana, IL 61801

*Corresponding author: hshan@illinois.edu

## Abstract

The multiphase flow of droplets is widespread, both at the industrial and the microscale, for both biological and non-biological applications alike. But the ensemble interactions of such systems are inherently nonlinear and complex, compounded by interfacial effects, making it a difficult many-body problem to understand, especially for theory. In comparison, the self-assembly dynamics of solid particles in flow have long been described and successfully exploited in the field of inertial microfluidics, where particle crystals can be realized from inertial forces and hydrodynamic interactions. Here, we report novel self-assembly dynamics of liquid drops in confined microfluidic channels that contrast starkly with the established paradigm of inertial microfluidics, which stipulates that higher inertia leads to better spatial ordering. Instead, we find that the most commonly used straight wall channel geometry not only fails to achieve regular spatial ordering for drops but actually exacerbates it with increasing inertia. Conversely, an asymmetric serpentine geometry is able to achieve long-range, periodic spatial ordering over length scales that are at least 3 orders of magnitude greater than the drop diameter, particularly at low inertia. Experimentally, we are able to decouple droplet generation from ordering, enabling independent variation of the additional parameters of number density, confinement, inertia, and surfactant concentration. We find the inertia-dependent emergence of preferred drop separations and show for the first time that Marangoni effects can influence the longitudinal ordering of multidrop arrays, with less surfactant being more detrimental at lower confinement. These results present a largely unexplored direction for inertial microfluidics but also show the potential of its unification with the versatile field of droplet microfluidics. In particular, the utility of passively restoring uniform drop spacing on-chip is a key requirement for the streamlined integration of incubation and drop-by-drop interrogation capabilities, which would greatly enhance the many droplet-based biological assays currently in use.

## I. Introduction

The field of inertial microfluidics [1–4] has seen growth in the past decade due to the potential for many applications that can make use of the passive, label-free, and high-throughput nature of inertial microfluidics. Examples include particle/cell sorting and enrichment [2,5–11], particle self-assembly [12,13], deterministic cell encapsulation [14,15] for overcoming Poisson loading, and micro-mixing [1,16–19]. Inertial focusing originates in 1961, when Segré and Silberberg observed particles focusing into an annulus under laminar flow in a circular pipe [20]. This discovery triggered subsequent studies to better understand the dynamics of inertial focusing [21–24], though they were under the context macroscopic flows, where dominance of fluid inertia was a given. In contrast to microfluidics, where flow rates are typically low and length scales are small, the prevailing thought had instead been that inertial forces and inertial effects were negligible [25]. This assumption holds true when the dimensionless Reynolds number, $Re = \rho UH/\mu$, is << 1. Here, $\rho$ is the mass density of the fluid, $U$ is the average fluid velocity, $H$ is the hydraulic diameter, and $\mu$ is the dynamic viscosity of the fluid. Yet, it is not difficult to achieve microfluidic flows with a Reynolds number of order 1 or higher [1], where the aforementioned inertial particle focusing can manifest. And indeed, this was demonstrated by Di Carlo et al. in 2007 [26], thus leading the way for inertial focusing to be harnessed at the microfluidic length scale for the manipulation and spatial ordering of solid particles and cells [2]. However, the potential of inertial ordering has not been similarly harnessed for liquid drops due to a deficient understanding of the different multiphase fluid dynamics governing the longitudinal or axial interactions between drops—interactions that we will show can give rise to previously undescribed self-assembly dynamics. Such dynamics can either result in ordered droplet crystals with long-range persistence or longitudinal disorder that positively correlates with fluid inertia against the established paradigm of inertial microfluidics.

Inertial microfluidic studies of particle ordering fall broadly under two particle types: solid particles and droplets. Solid particles—commonly beads—are the most extensively studied, being the simplest go-to model and having direct applicability to cells [2]. Although cells are deformable, they have been more comparable to solid particles than to liquid drops due to several distinguishing traits. One is the confinement, defined as $a/H$, where $a$ is the particle diameter. Confinement significantly influences the flow and hydrodynamic forces surrounding both solid particles [27] and drops [28–30]. For beads and cells, confinement is low in practice (typically ≤ 0.4) to avoid clogging [26,31]. In contrast, microfluidic drops are typically made at high confinement to minimize wasteful oil use. Another trait is that both cells and solid particles rotate in flow, whereas drops experience internal fluid circulation instead [32,33], which can affect drop dynamics via surfactants. Indeed, internal fluid circulation is coupled to perhaps the most defining traits for drops: the fluid viscosity ratio [34,35] and the presence of interfacial/Marangoni effects [36–42], both of which differentiate drop deformability from that of cells, which is instead governed by complex viscoelastic cellular components. These differences demonstrate that drops are a distinct system requiring separate consideration, which has been duly recognized in the form of dedicated studies investigating the lateral dynamics of drops [35,41–48]. These studies have shed light on the deformation-induced lift force and the unique impact of surfactants and Marangoni effects on lateral dynamics. While the lateral aspect is important, it can be analyzed in the dilute limit and neglects the ensemble behavior and self-organization which can arise from the many-body drop interactions that are necessarily part of longitudinal ordering and real systems. The longitudinal dimension, however, has remained largely unexplored for drops, not just experimentally, but also computationally and theoretically due to the immense difficulties of a many-body problem with multifaceted nonlinear interactions, thus limiting such studies to more tractable pair-wise interactions [49,50].

Investigation into the longitudinal inertial ordering of solid particles has been more extensive, and have documented hydrodynamic forces acting between them [12,51,52]. Research into these particle-particle interactions as well as the stability of particle trains have spanned both Newtonian fluids [53–59] and non-Newtonian ones [13,31,60]. While substantial and of benefit to the wide use of particles in microfluidics, these efforts have primarily been phenomenological, and have not yet culminated in a general theoretical framework capable of robustly predicting interparticle spacings. Yet, a rivaling body of work for drops achieving at least a similar level of understanding is—to the best of our knowledge—severely lacking, even though the application of drops in microfluidics has certainly not been [61,62].

Droplet microfluidics has enabled a myriad of biotechnologies [63,64], such as single-cell [65] and single-virus [66] sequencing. Despite its immense potential, drop microfluidics is mostly limited to single-step reactions per chip [65–69] and performing multistep reactions in drops still poses practical challenges. Most biological assays involve repeated incubation and reagent addition steps. Unfortunately, on-chip incubation often disturbs even spacing of drops, making it difficult to manipulate drops downstream [70]. Disruption of uniform drop spacing is particularly disadvantageous for reagent addition, picoinjection [71], and sorting. As a result, multistep procedures tend to be performed off-chip, followed by reinjection of drops into a separate device. However, reinjection often leads to drop merging and loss of material during flow rate optimization. Oil removal modules have been developed to restore regular spacing [72], but this requires delicate flow balancing, preventing simple and effective integration of on-chip incubation and drop manipulation. Therefore, a better understanding of the conditions for passively achieving the longitudinal ordering of drops could streamline on-chip integration and improve throughput by reducing the number of off-chip processing steps. Such conditions must also include low flow rates, which are necessary to facilitate drop-by-drop screening [73,74] and to ensure sufficient on-chip incubation times. However, the conventional wisdom of inertial microfluidics stipulates high **mL·hr$^{-1}$** flow rates to achieve inertial ordering for solid particles. We will show that the practical high confinement of microfluidic drops enables the use of lower flow rates, with effectual particle Reynolds numbers $\boldsymbol{R_p}$ < 1, where $\boldsymbol{R_p = Re(a/H)^2}$. The dynamics presented here could not only remedy the aforementioned deficiencies but also spur the new subfield of inertial droplet microfluidics.

In this study, we explore how the parameters of channel geometry, number density, confinement, inertia, and Marangoni effects influence the longitudinal inertial ordering of drops. In particular, we employ high confinement ($\boldsymbol{a/H}$ = 0.8 and 0.64) and particle Reynolds numbers ranging from $\boldsymbol{O}$(0.1) to $\boldsymbol{O}$(1). Surprisingly, our comparison of the traditional straight wall channel geometry to the asymmetric serpentine channel geometry [8,26] shows that not only is the latter capable of establishing persistent periodic droplet arrays, but the former actually exacerbates spacing heterogeneity—especially at higher inertia—which is counter to the paradigm of inertial microfluidics. Through independent variation of the experimental parameters, we identified conditions that can enable long-range ordering of coherent droplet arrays on length scales that are at least three orders of magnitude greater than the drop diameter. Furthermore, we experimentally demonstrate for the first time that Marangoni effects can directly influence longitudinal inertial ordering. Overall, this study presents novel self-assembly dynamics of an interacting, three-dimensional, and many-body dispersed flow while also demonstrating the practical feasibility of achieving long-range longitudinal ordering for drops at very low flow rates (**60 μL·hr$^{-1}$**), which are well-suited to droplet microfluidic applications.

## II. Experimental Design

For spacing measurements, fluorescent drops are first made separately using 2% surfactant (see Appendix) and then reinjected into inertial ordering devices. By reinjecting the drops instead of inertially ordering them on the chip in which they are made, drop size, flow rate, and drop number density can all be decoupled and varied *independently*, thus overcoming a common experimental limitation inherent to most studies investigating hydrodynamic phenomenon with microfluidic drops. These reinjected drops are then respaced using oil but are immediately randomized so as to disrupt the otherwise even spacing that the oil would create (see Fig. 1A-C). Drop coalescence during reinjection is sensitive to various factors, but can be minimized through elimination of static and slow, gentle pipetting. However, if coalescence is high, data is not taken. Limited coalescence is acceptable (approximately 2-3 per 1500 drops), and they are deleted from the data set.

Drops flowing through the main ordering channel are measured optically using a laser line to detect the fluorescent drops at set positions along the channel in 1 cm increments. Fluorescence is detected at a photomultiplier tube (PMT) and converted to voltage peaks (see

Supplementary Fig. S1 for example PMT trace). Data is taken using a custom LabVIEW™ program that records peak size and time between peaks. This method allows rapid automated recording of drops without resorting to more involved image analysis. The time values measured are used to measure the edge-to-edge distance between drops and to directly evaluate the dispersion of the droplet spacing distribution, which is characterized by the dimensionless coefficient of variation $CoV$:

$$CoV = \frac{S}{\langle \mathbf{X} \rangle}, \qquad (1)$$

where $S$ is the standard deviation and $\langle \mathbf{X} \rangle$ is the mean. A lower $CoV$ represents a more homogeneously spaced and well-ordered droplet array.

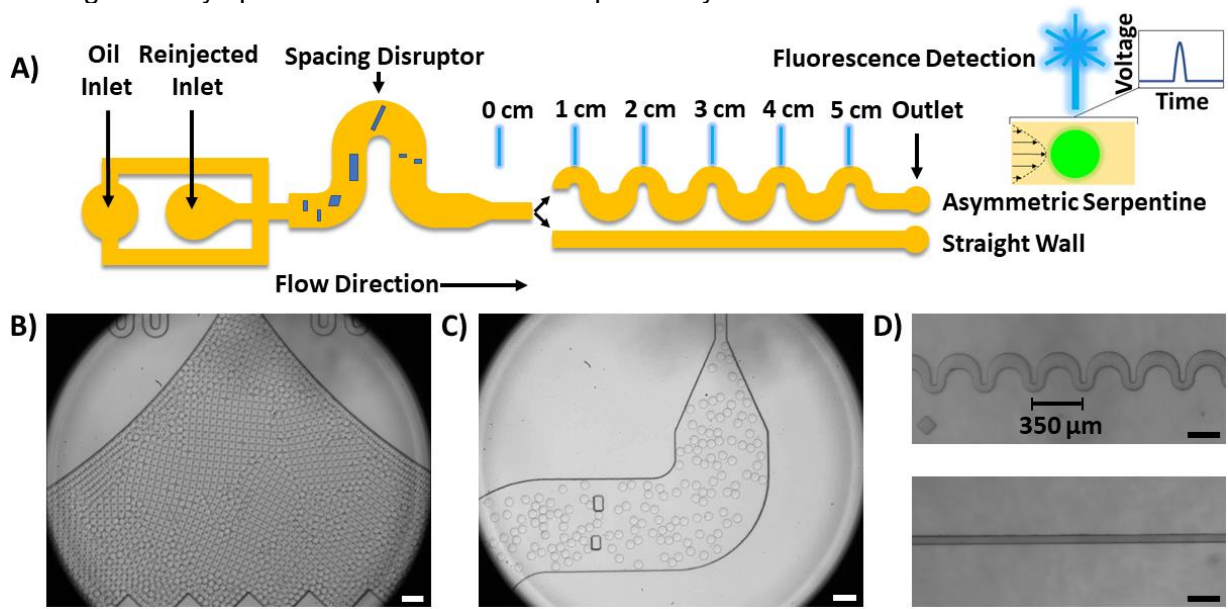

**Figure 1.** Overview of device used for spacing measurements with reinjected drops. (A) Device diagram. Reinjected drops are close-packed and then respaced with fluorinated oil. To avoid creating even spacing, drops pass through a spacing disruptor region where obstructions assist with randomizing drops. The ordering portion of the device (asymmetric serpentine or straight geometry) follows this spacing disruptor region (continuing toward the right) and extends **5 cm** along the chip. Spacing measurements are made at **1 cm** increments by fluorescence detection. (B) Example of close-packed drops after reinjection. (C) Image showing drops being scrambled in the spacing disruptor region. (D) Top image shows asymmetric serpentine geometry, with channel height of **50 μm** and radii of curvature of **25** and **75 μm** for the smaller curve and **100** and **200 μm** for the larger curve. The periodicity is **350 μm**. Bottom image shows the straight wall geometry, with channel height and width at **50 μm**. For the asymmetric geometry, the hydraulic diameter is calculated using the narrowest part of the curve, as that is where the spacing measurements are made with the laser. Scale bars (white) in B and C are **100 μm**. Scale bars (black) in D are **200 μm**.

## III. Results and Discussion

### A. Longitudinal Inertial Ordering of Droplets Depends on Channel Geometry

Especially for the practical aspect of identifying suitable conditions for the longitudinal inertial ordering of drops, channel geometry is an important consideration. We tested two geometries: the standard straight wall and the asymmetric serpentine (Fig. 1A and 1D), both with typical rectangular cross-sections. Due to Dean flow [75], serpentine geometries have been shown to achieve lateral focusing within shorter distances and at lower $R_p$ than the straight wall geometry for solid particles at low confinement [26], with the asymmetric variant biasing particles to a single stream instead of two. Lateral ordering was considered as it appears that longitudinal ordering may possibly only follow when the former is achieved first [51]. Different from previous works, our study with high confinement droplets show that the traditional straight wall geometry not only failed to improve spacing uniformity, but actually worsened it (Fig. 2B). In comparison, the asymmetric serpentine geometry was able to achieve stable long-range, longitudinal ordering of continuous single-file drop trains over distances that are at least 3 orders of magnitude greater than the drop diameter (Fig. 2A). This long range stands in stark contrast to the short 10-15 particle trains reported previously [26], and only represents a lower limit on the range because the ordering channel was not made longer. The histogram of droplet spacing at the **5 cm** mark in the asymmetric serpentine channel shows significant narrowing compared to the initial distribution (Fig. 2C, total flow rate: **60 µL hr⁻¹**, $R_p$: 0.139). Fig. 2A and 2G together show that the spacing improvement could be achieved quickly (within **3 cm**, contour length ~**6 cm**), and is stable afterwards. Fig. 2D and 2H similarly compare the distributions of drop spacing in the straight wall geometry at $R_p$ = 0.139, but the geometry-specific difference is stark; at the downstream position, the shape of the distribution does not evolve toward spacing uniformity and instead shows an enrichment of small spacing counts as well as more frequent large spacings (> **400 µm**). The spacing distribution also evolves very differently with inertia for each geometry. In the asymmetric serpentine (Fig. 2E), the shape and peak position remain the same, but in the straight wall (Fig. 2F), the counts for large spacings increase further with inertia, along with a significant increase in counts for even smaller drop spacings than before. Supplementary Fig. S2 also shows this inertia-dependent progression for the two intermediate flow rates. These results are counter to the conventional wisdom of inertial microfluidics developed predominantly from studying solid particles in straight wall channels, whereby higher inertia tends to produce stronger ordering.

The increasing polarization of large and small drop spacings in the straight wall geometry underlies the increasing $CoV$ plotted in Fig. 2B. This polarization likely indicates the emergence of at least 2 fixed points for drop separation in the straight wall geometry at high confinement. These fixed points appear to be stable as they act as attractors that increase the counts for both very large and small drop spacings. On the contrary, drop separation in the asymmetric serpentine appears to exhibit only one fixed point, which is stable. The likely presence of two fixed points in the straight wall geometry qualitatively disagrees with a recent numerical study looking at a limited case of a pair of non-spherical drops that are perfectly centered in a straight channel [50]. This study identified only a single stable fixed point at similarly high confinement, though it used viscosity and mass density ratios of 1, and Marangoni effects were ignored. Our study, however, reflects the practical conditions that are commonly used for droplet microfluidics. The aid of further numerical studies will likely be necessary to fully elucidate this matter.

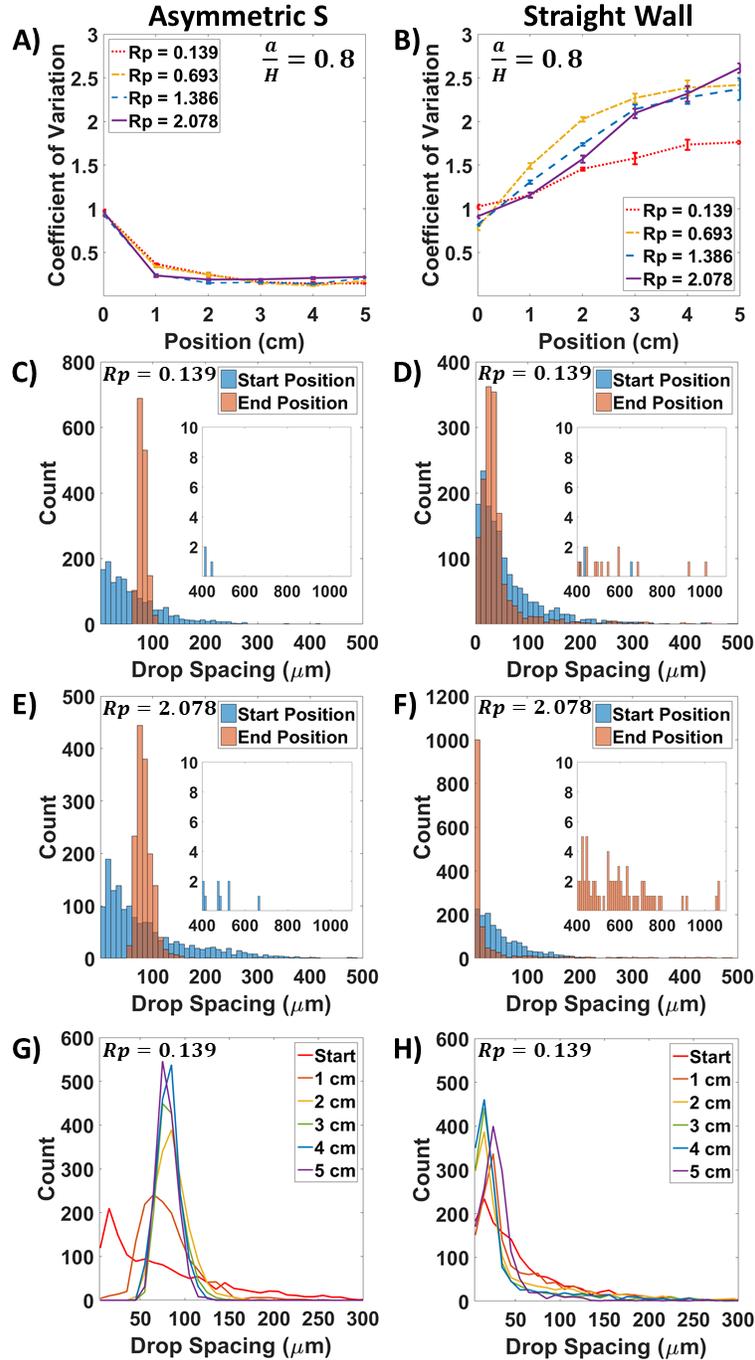

**Figure 2.** Comparison of channel geometries. Both use a droplet number density of ~**3700 drops·μL$^{-1}$** (denoted medium number density) and confinement $a/H$ = 0.8 ($a$ = 40 μm). $CoV$ averages and error bars (standard deviations) are plotted in A and B, with each point coming from 3 to 4 trials, and for each trial $N$ = **1500** drops. $N$ = **1500** drops for C-F and for each histogram at each channel position in G and H. Bin sizes for histograms are **10 μm**. Total flow rates corresponding to $R_p$ values (low to high) are **60**, **300**, **600**, and **900 μL·hr$^{-1}$**, or average velocities of approximately **0.667, 3.33, 6.67**, and **10 cm·s$^{-1}$**, respectively. (A) Asymmetric serpentine channel geometry shows decreasing $CoV$ with distance traveled for all tested flow rates. (B) Straight wall geometry shows increasing $CoV$, indicating worsening spacing distribution with distance traveled. (C) Example histogram of spacing distribution in the asymmetric serpentine geometry for the lowest $R_p$ = 0.139. Start position and end position refer to data plotted at the **0 cm** and **5 cm** positions, respectively. Inset is a zoomed in view on the counts for the largest spacings. (D) Example histogram of lowest $R_p$ case in the straight wall geometry. Here, inset shows that there are more occurrences of these large spacings for drops at the end position than at the start. (E) and (F) Histograms of the asymmetric and straight wall geometries at $R_p$ = 2.078, respectively. Note the more polarized spacing distribution between D and F. (G) and (H) Histograms of spacing distribution as a function of position along the channel. Stable ordering is achieved in the asymmetric serpentine.

The underlying ordering of drops in each geometry is further analyzed through plotting the autocorrelation function (ACF) as a function of lag, as shown in Fig. 3. The autocorrelation $A_k$, for lag $k$, is defined as:

*( 2 )*

$$A_k = \frac{V_k}{V_0},$$

where $V_0$ is the variance of the base time series $\mathbf{X}$ of length $T$, and $V_k$ is the covariance of $\mathbf{X}$ with its lagged version, given by:

*( 3 )*

$$V_k = \frac{1}{T}\sum_{t=1}^{T-k}(X_t - \langle \mathbf{X} \rangle)(X_{t+k} - \langle \mathbf{X} \rangle),$$

As expected, the spacing disruptor randomizes the initial drop spatial distribution (Fig. 3A and 3B). The ACF trace generated from $N \approx$ **1500** drops shows the periodic ordering of drops in the asymmetric serpentine, evidenced by the peaks at multiples of **80 μm** (Fig. 3C) and the steady state images (Fig. 3G). This ordering is also stable over time, which is shown through the ACF of $N$ = **6000** drops combined from multiple data sets (Supplementary Fig. S3). The periodic nature of the ACF directly relates to the low $CoV$ (~0.15) shown in Fig. 2A. However, higher inertia leads to slightly poorer ordering in the asymmetric serpentine, resulting in a faster decay in the ACF (Fig. 3E) and a slightly higher $CoV$ (~0.2, Fig. 2A). In contrast, the ACF trace for drops in the straight wall geometry does not show periodic spacing. At the lowest inertia (Fig. 3D), the ACF shows a secondary peak away from 0, which reflects short-range correlations from neighbor-to-neighbor hydrodynamic interactions. There may be a degree of very localized ordering at the lowest inertia, which is supported by a lesser $CoV$ rise shown in Fig. 2B and the peak position being away from 0 in Fig. 2D. At higher inertia (Fig. 3F), the ACF does not initially decay below 0 and then rise to a secondary peak, but instead the decay toward statistical insignificance is fairly monotonic, reflecting the change in ordering dynamics that resulted in greater spacing polarization (Fig. 2F). An example of the visual difference between the low and high inertia cases is shown in Fig. 3H, where drops tend to self-organize into clusters at higher inertia. Despite the lack of uniform spacing being established for drops in the straight wall geometry, the ACFs show that the organization is not random (as opposed to Fig. 3B). This reflects undescribed hydrodynamic interactions for high confinement drops in the conventional straight wall geometry—hydrodynamic interactions that are very much different from those in the asymmetric serpentine geometry.

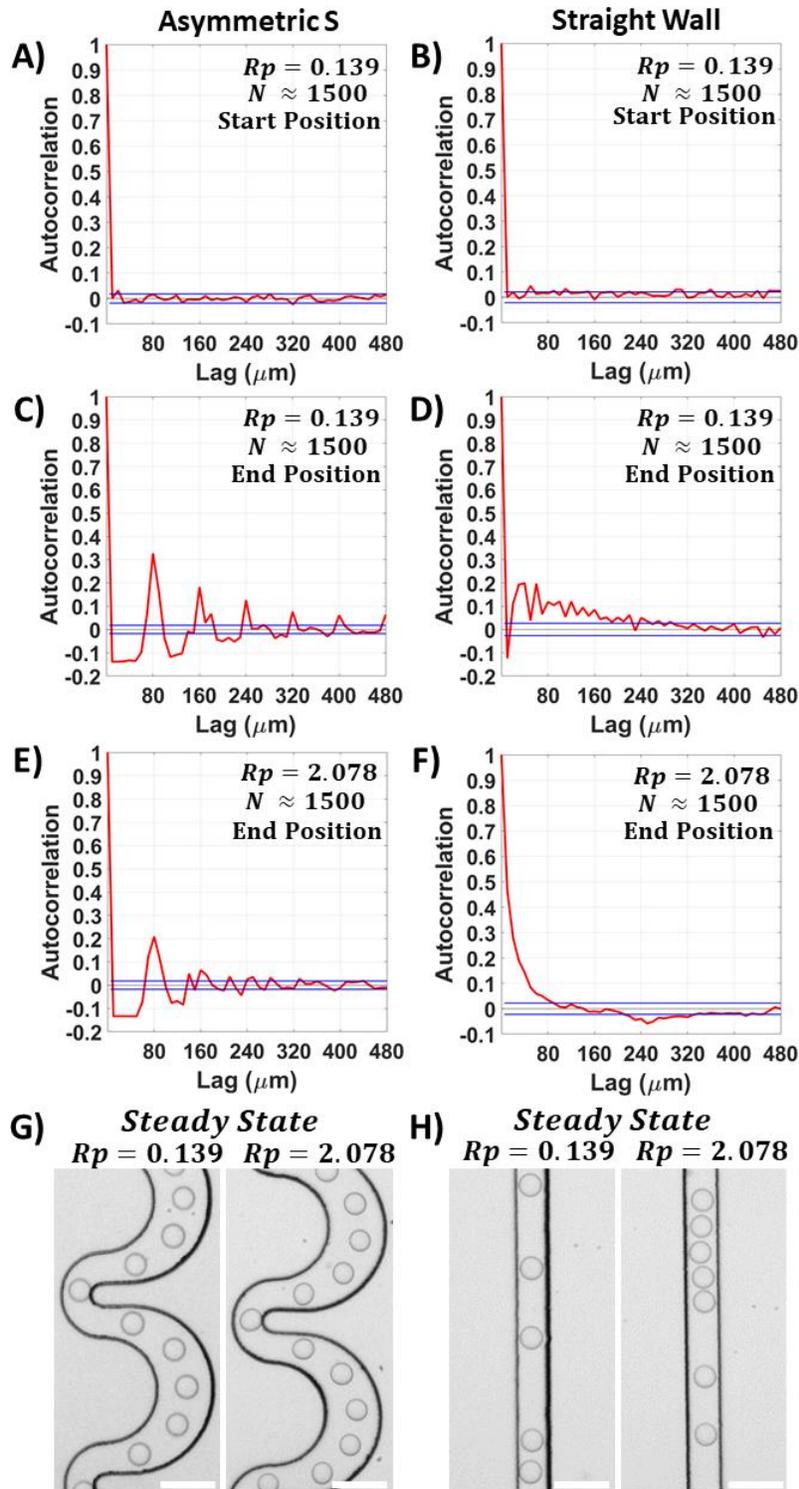

**Figure 3.** The autocorrelation functions for both geometries and $a/H = 0.8$ are plotted as a function of lag. Horizontal blue lines represent 95% confidence intervals. (A) and (B) The ACFs at the start position for the asymmetric serpentine and straight wall geometries, respectively, showing a randomized spacing, as intended. (C) and (D) The ACFs at the end position, or **5 cm** position, in the asymmetric serpentine and straight wall geometries, respectively. (E) and (F) ACFs for the asymmetric serpentine and straight wall geometries,

respectively, at the highest flow rate used. (G) and (H) These images are examples of the downstream steady state of the drops. They are well-spaced in the asymmetric geometry. In the straight wall geometry, drops are not well-spaced but still have inertia-dependent organization. Scale bars are **100 µm**.

The temporal, and by extension spatial, persistence of the ordered droplet array and the range of the hydrodynamic interactions between drops can be characterized through the Hurst exponent $h$ via rescaled range analysis [76], where $h$ is a measure of the decay of the ACF and an indication of long-range dependence.

(4)
$$\mathbb{E}\left[\frac{R^i(n)}{S^i(n)}\right] = Cn^h,$$

Here, $\mathbb{E}$ denotes the expectation value over $i$, $C$ is a constant, $R^i(n)$ is the rescaled range for the $i^{th}$ subset of size $n$, and $S^i(n)$ is the corresponding standard deviation. For a time series **X** of length $T$ (~1500), $n$ is chosen to be $= T, T/2, T/4, \ldots, T/128$. For each $n$, there are $i = 1 \ldots m$ subsets, where $m = T/n$, and $R^i(n)$ is defined as:

(5)
$$R^i(n) = \max[\mathbf{Z}^i(n)] - \min[\mathbf{Z}^i(n)],$$

where $\mathbf{Z}^i(n)$ is the cumulative deviation series given by:

(6)
$$\mathbf{Z}^i(n) = \left(\sum_{j=1}^{1} Y_j^i(n), \ldots, \sum_{j=1}^{n} Y_j^i(n)\right),$$

and $\mathbf{Y}^i(n)$ is the $i^{th}$ mean centered series:

(7)
$$\mathbf{Y}^i(n) = \mathbf{X}^i(n) - \langle\mathbf{X}^i(n)\rangle,$$

The Hurst exponent $h$ is found from the slope of the linear fit after log-transforming Eq. 4:

(8)
$$\log \mathbb{E}\left[\frac{R^i(n)}{S^i(n)}\right] = h \log n + C',$$

Shown in Fig. 4 are representative time series (i) of the drop spacings, corresponding to the ACF's plotted in Fig. 3. Accompanying each is a log-log plot (ii) of the rescaled range analysis, with the slopes of the linear fits given as $h$. For the drops in the asymmetric serpentine at $R_p$ = 0.139 (Fig. 4A), there is a high $h$ of 0.84(4), indicating strong long-range dependence and in agreement with the low $CoV$ in Fig. 2A and the periodic ACF in Fig. 3C. Similarly, $h$ is lower but still > 0.5 at $R_p$ = 2.078 due to the slightly poorer ordering at high inertia (Fig. 4B). For drops in the straight wall channel, $h$ = 0.52(2) at $R_p$ = 0.139 (Fig. 4C), indicating a lack of long-range dependence. However, an $h$ of 0.5 does not preclude the existence of short-range correlations, which is consistent with Fig. 3D. For the drops at $R_p$ = 2.078 in the straight wall channel (Fig. 4D), the fluctuations in the time series are much more frequent and greater in amplitude. Correspondingly, $h$ = 0.36(2), which is indicative of the fluctuations between high and low values. These fluctuations also reflect the increased $CoV$ (Fig. 2B) as the spacing heterogeneity reflects a greater dispersion in the distribution. In addition, Supplementary Fig. S4 shows a representative rescaled range analysis on the randomized initial drop distribution measured at the start position, confirming a lack of long-range correlations, $h$ = 0.48(2), as expected.

Although ordering for solid particles is known to be achievable in both channel geometries, our results show that a serpentine geometry is necessary for drops, suggesting a critical role played by the nonlinear Dean flow. Previous studies on neutrally buoyant, solid particles at low confinement showed that the Dean drag force $F_D$, which scales as:

$$F_D \sim \frac{\rho U^2 a H^2}{R},$$

(9)

where $R$ is the radius of curvature of the channel, induces mixing and can help particles reach equilibrium faster through quicker sampling of their lateral positions [26]. These studies show that particle ordering is most efficient when the ratio between the inertial lift force and the Dean drag force is ~1. If the ratio is >> 1, then inertial lift forces are dominant, and no advantage will be gained from the Dean flow. If the ratio is << 1, the Dean drag force becomes disruptive. As guidance [17], Di Carlo suggests a ratio, given by $2Ra^2/H^3$, to be at least > ~0.08. Here, the ratio is of order 1 for the smallest average radius of curvature used (**50 μm**) and a drop diameter of **40 μm**, which is ideal [26]. While this is perhaps supportive of the asymmetric serpentine results as the ratio should be satisfied at all $R_p$ used, it is necessary to acknowledge the fact that these are high confinement drops, not solid particles, and deformation, buoyancy, as well as Marangoni effects may influence the criterion for drop ordering in undescribed ways. To better understand the flow topology and hydrodynamic forces among deformable drops in serpentine geometries, numerical analysis of such systems will likely be required. Recent computational efforts have acknowledged the need to better understand flows in serpentine geometries, further investigating both the base fluid behavior in these channels [77] as well as the forces imposed upon small solid particles [78].

The characteristic of buoyancy is also at play in our system due to the different mass densities of the aqueous dispersed phase and the oil continuous phase, which is typical in droplet microfluidics and most systems with two immiscible fluids. Density matching is commonly and easily done for solid particles by slightly adjusting the density of the aqueous carrier fluid. However, this is impractical for drop systems cannot be done without appreciably changing the viscosity ratio and therefore fundamentally altering the system, especially here, where the continuous phase ~1.6x denser than the dispersed phase. We employed CsCl salt to minimize changes to viscosity while increasing the dispersed phase density, but even so, the viscosity was appreciably altered (see Supplementary Fig. S5, which shows visible differences in drop deformation). On account of the viscosity change, as well as observations that ordering was not prevented in the asymmetric serpentine nor enabled in the straight wall channel after density matching, this was not pursued further. However, we also sought to verify if the inherent buoyancy presented a problem if the magnitude was large enough to force drops to rise up and engage in frictious contact with the upper channel wall. A comparative calculation was done using the net upward force $F_{B,net}$ on the drop due to the buoyant force and the weight of the drop vs the net empirical lift force $F_{L,empirical}$ on the drop [41], which is pointed toward the channel center. $F_{B,net}$ is given by:

(10)
$$F_{B,net} = \left(\frac{4\pi}{3}\right) r^3 g (\rho_{out} - \rho_{in}),$$

where $r$ is the drop radius, $g$ is the gravitational acceleration, $\rho_{out}$ is the mass density of the continuous phase, and $\rho_{in}$ is the mass density of the dispersed phase. $F_{L,empirical}$ is given by:

(11)
$$F_{L,empirical} = C_L \mu U r \left(\frac{r}{H}\right)^3 \left(\frac{d}{H}\right),$$

where $C_L$ is the empirical lift coefficient, which the authors recommend to be 500 for general estimates, and $d$ is the distance of the drop center from the channel center. For numerical details, see Supplementary Material. But the calculation shows that $F_{L,empirical} > F_{B,net}$ even at the lowest flow rate used and for both drop sizes. Thus, buoyancy here is insufficient to push drops up against the top channel wall, avoiding the complication of friction effects influencing drop behavior.

     With contrasting duality, we demonstrated aperiodic self-assembly dynamics against the inertial microfluidic paradigm while also showing the longitudinal ordering of droplet crystal arrays on length scales that are at least three orders of magnitude greater than the drop diameter. Furthering this dichotomy, the lack of ordering for the former worsened with higher inertia whereas the spatial uniformity for the latter was best at the lowest $R_p$ of 0.139 (**60 μL·hr$^{-1}$**), also against the norm of high **mL·hr$^{-1}$** flow rates required for solid particles. Practically, this long-range ordering ($h > 0.5$) is critical for the consistent, deterministic manipulation of drops, and the low flow rates are necessary for accurate drop-by-drop screening, sorting, and picoinjection. The passive inertial ordering of drops shown here greatly extends the potential of droplet microfluidics by enabling a streamlined integration of on-chip incubation and drop manipulation functions.

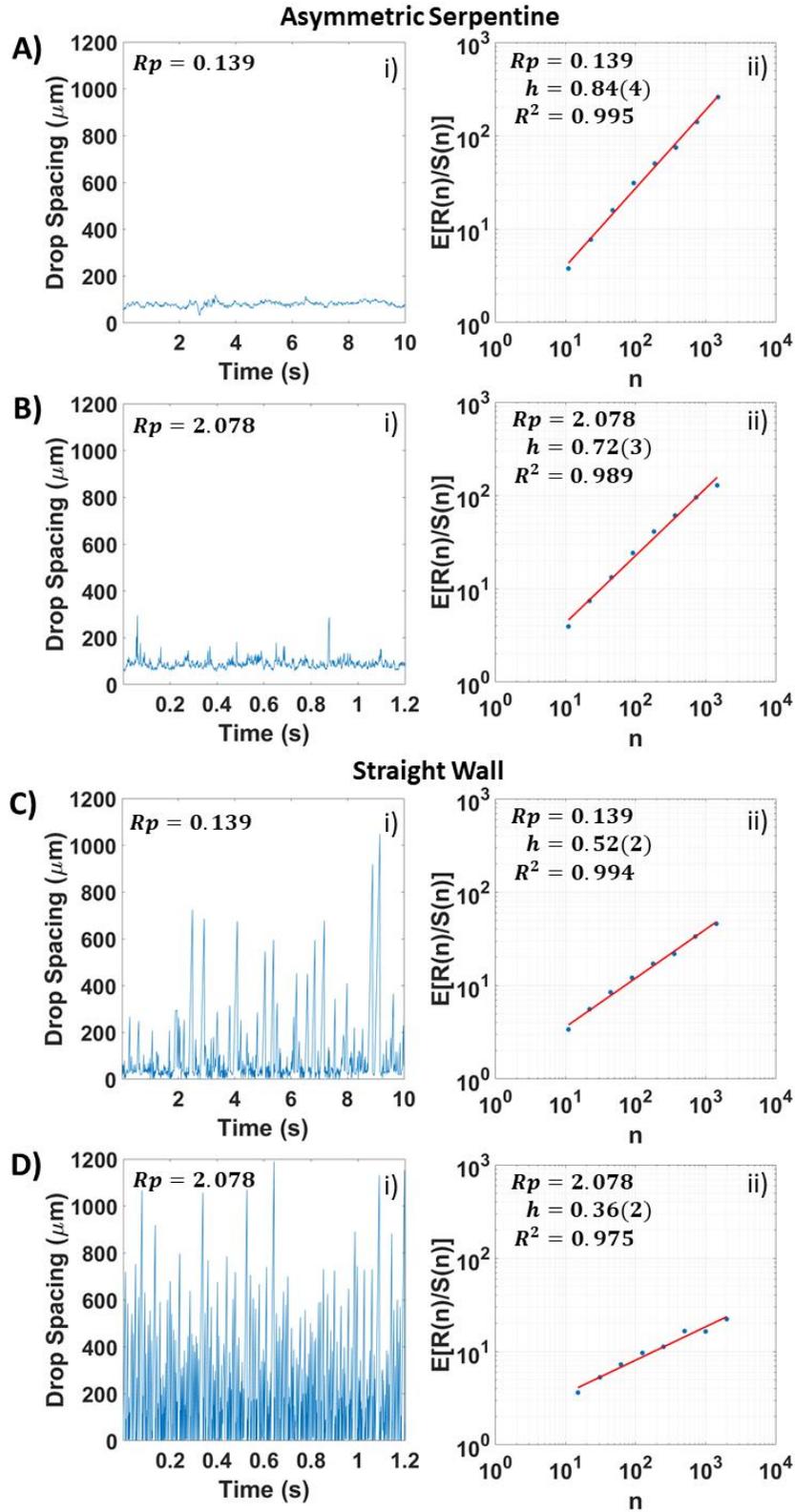

**Figure 4.** Example time series of observed drop spacings at the highest and lowest $R_p$ in both channel geometries at the **5 cm** position and for $a/H$ = 0.8 are shown in (i). In (ii) are the log-log plots of the results from the rescaled range analysis of the corresponding time series. The slope of the linear fit (standard error in parentheses) on log-log scale gives the Hurst exponent $h$, a measure of long-range dependency. R-squared for the fits are also shown. (A) and (B) Asymmetric serpentine geometry. (C) and (D) Straight wall geometry.

## B. Number Density Alters Hydrodynamic Interaction Strength and Ordering

For the longitudinal interactions themselves, the parameter that most directly modulates their strength, which scales inversely with the distance between the drops, is the drop number density. Shown in Fig. 5A is a diagram of the base forces experienced by the drops in both geometries—discounting interfacial effects—including the known lateral lift forces, hydrodynamic repulsion, as well as the net effect of buoyancy. These forces have been shown experimentally for solid particles in a straight wall channel [17,51,52]. Similar to the solid particle case, the primary longitudinal force is a hydrodynamic repulsion between the drops, which arises from the reversing streamlines that develop between them [49]. Fig. 5B directly visualizes the presence of the repulsive force in the asymmetric serpentine geometry, which also induces Dean flow. The left most image shows a continuous train of drops, while the middle and right images show the end of the train later in time, when a disruption to the continuous train is introduced. The three drops framed in yellow, blue, and red are the last drops in the train. It is evident that the spacing at the end is not only greater than the spacing in the continuous drop train ($d_0$), but the spacing actually increases between the last and second last drops ($d_2$) compared to the second and third last drops ($d_1$). This observation is explained by the mutual repulsion between them and the boundary condition of not having a neighbor at the end. As a result, the last drop feels a repulsion in only one direction, which moves it further from the drop in front. The increased distance then weakens their mutual repulsion, allowing the second last drop to also move further away from its forward neighbor ($d_0 < d_1 < d_2$).

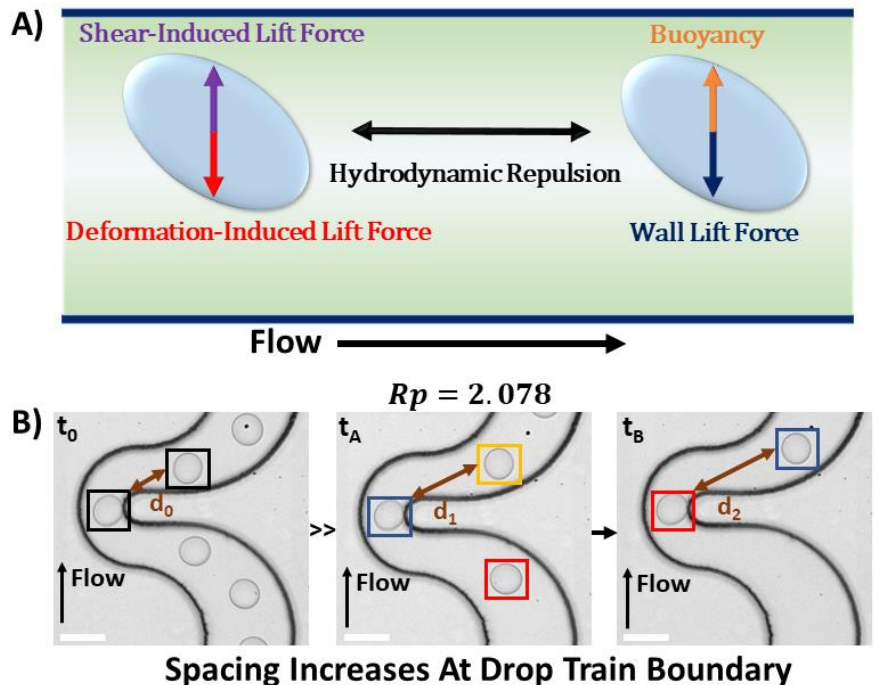

**Figure 5.** This figure illustrates the base forces common to both channel geometries. Dean drag forces (not shown) are present in the asymmetric serpentine geometry only, and are perpendicular to the primary flow direction. (A) The lateral lift forces are well known, and are present for lone particles. The longitudinal hydrodynamic repulsion is the primary force directly

influencing self-ordering along the direction of flow and is affected by the number density. Note that buoyancy here refers to the net force on the drop as a result of the buoyant force and the weight of the drop; in this study, the dispersed aqueous phase has a lower mass density than the continuous oil phase. (B) These images show that even in the more complex case of the asymmetric serpentine geometry where Dean flow is present, hydrodynamic repulsion still manifests. The image on the far left ($t_0$) shows a steady train of drops, with $d_0$ depicting the distance between two drops. After some time ($t_A$), there is a disruption to the drop train that is introduced. The three drops framed in yellow, blue, and red are the three last drops in the train. Note that the spacing between the yellow-framed drop and the blue-framed drop ($d_1$), is greater than $d_0$ but smaller than $d_2$, as depicted in the right image ($t_B$). Scale bars are **50 μm**.

Note that we use number density and not volume fraction because the former allows direct comparison between different sized drops. In other words, if drops in a single-file train have the same number density, their center-to-center distance is the same regardless of drop size. In contrast, the same volume fraction would result in approximately double the number density for **32 μm** drops vs **40 μm** drops (volume ratio ≈ 0.5), which implies a shorter mean drop-to-drop distance for **32 μm** drops. The number density, which is defined by the **drops·μL$^{-1}$**, is estimated using the aqueous flow rate, the total flow rate, and the volume of an individual drop. This estimate assumes that drops are close-packed during reinjection. The close-packing of equal solid spheres implies that the spheres themselves constitute approximately 74% of their occupied volume. Given that the drops are deformable, the number density in flow could be slightly higher. The number densities we have denoted as low, medium, and high are estimated to be approximately ~**1900**, ~**3700**, and ~**7400 drops·μL$^{-1}$**, respectively. Using these number density approximations, they are equivalent—inside an idealized infinite **50 μm** x **50 μm** straight wall channel where the drops are uniformly spaced—to a length per drop of approximately ~**210**, ~**110**, and ~**50 μm**, respectively.

Fig. 6A and 6B summarize the effects of number density as well as inertia on the ability of their respective geometries to affect ordering. Here, the degree of drop ordering is represented using a $CoV$ ratio, which is the ratio of the $CoV$ at the **5 cm** position to the **0 cm** position. This value measures the final improvement of spacing distribution. These results show that number density is critical to influencing the degree of improvement and the inertial dependence. Only the asymmetric serpentine geometry allows for meaningful spacing improvement ($CoV$ ratio notably < 1), with the medium number density condition, previously presented in Fig. 2A, showing the best reduction in $CoV$ and having insignificant dependence on $R_p$. In the high number density case, interactions are stronger, but it also means greater crowding and less space to rearrange out of nonuniform starting configurations. The disruptive effect of overcrowding has also been observed for solid particle trains [53,54] when suspensions are too concentrated. Unsurprisingly, it represents the worst number density for ordering (see Supplementary Fig. S6 for detailed $CoV$ data as a function of channel position). At the low number density, there is more space for drop movement, but longitudinal interactions are also weakened, and $CoV$ ratios are predominantly higher than those at the medium number density. Moreover, $R_p$ dependence emerges in this case with better ordering at lower $R_p$ than at higher $R_p$. The best $CoV$ ratio achieved in the low number density is still not as good as that in the medium number density case, possibly due to weakened longitudinal interactions. It is worth noting that an increase in average drop-to-drop distance does not prevent the drops from individually achieving lateral focusing. Therefore, the fact that the $CoV$ increases with $R_p$ much more strongly at low number density than at medium number density suggests the sole influence of longitudinal interactions in countering a disruptive effect that strengthens with higher inertia.

Different from the asymmetric serpentine geometry, the straight wall geometry significantly exacerbates drop spacing in all conditions ($CoV$ ratio > 1, Fig. 6B). The effect of weakened longitudinal interactions due to lower number density is also seen in the straight wall geometry (more detail in Supplementary Fig. S6D). Despite the continued trend of increasing $CoV$, it does not increase as much overall as compared to the high (Supplementary Fig. S6C) and medium (Fig. 2B) number density cases.

Our results emphasize the importance of drop number density for longitudinal ordering. Although it is evident that the straight wall geometry systematically fails to improve longitudinal spacing uniformity, the asymmetric serpentine geometry does not just automatically guarantee good ordering. Instead, the number density dictates the ordering capability (Fig. 6A), with no meaningful improvement at high number density. Moreover, inertial dependence only emerges for the low number density condition in the asymmetric serpentine. These results reaffirm that both the particle concentration—which sets the strength of longitudinal interactions—and the accompanying crowding effects play crucial roles in governing the dynamics of multidrop trains.

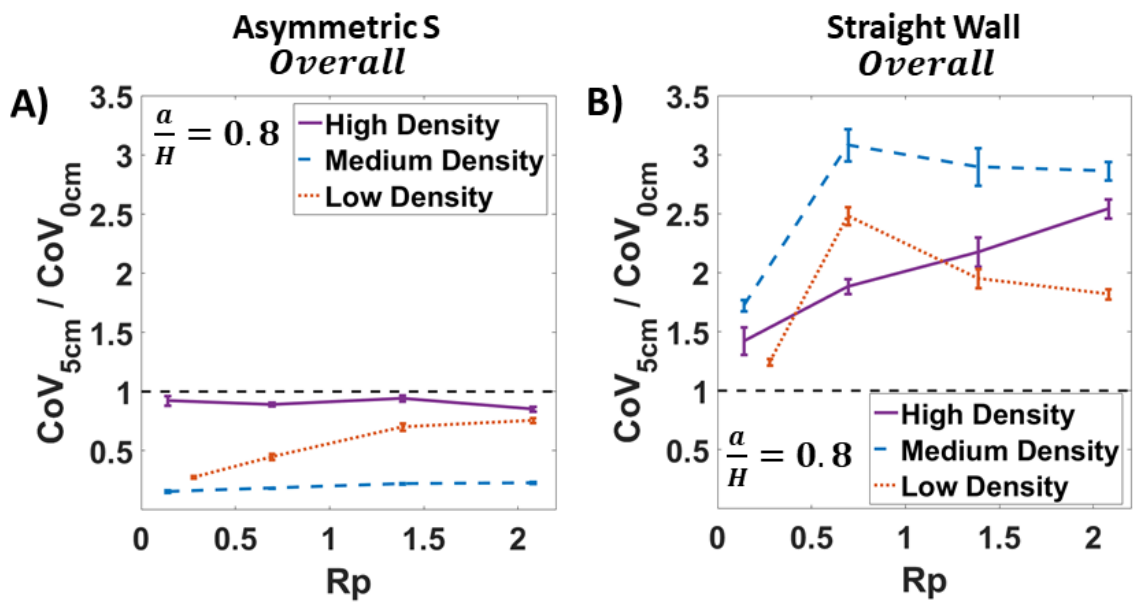

**Figure 6.** Spacing results demonstrating the effect of number density on the previously observed trends in the asymmetric serpentine and straight wall geometries. $CoV$'s are originally computed in the same manner described in Fig. 2, with $N \approx$ **4500-6000** drops per data point. Confinement is $a/H$ = 0.8 for all cases. Flow rates corresponding to the three highest $R_p$ values are the same as before, but the 0.277 value in the low number density case corresponds to a flow rate of **120 μL·hr$^{-1}$**. $CoV$ data as a function of channel position is shown in Supplementary Fig. S5. (A) and (B) Asymmetric serpentine and straight wall geometries at all 3 number densities, respectively. Data plotted represents the ratio of the $CoV$ at the **5 cm** channel position to the $CoV$ at the **0 cm** position.

### C. Confinement Effects on Longitudinal Ordering, Fixed Points, and Inertial Dependence

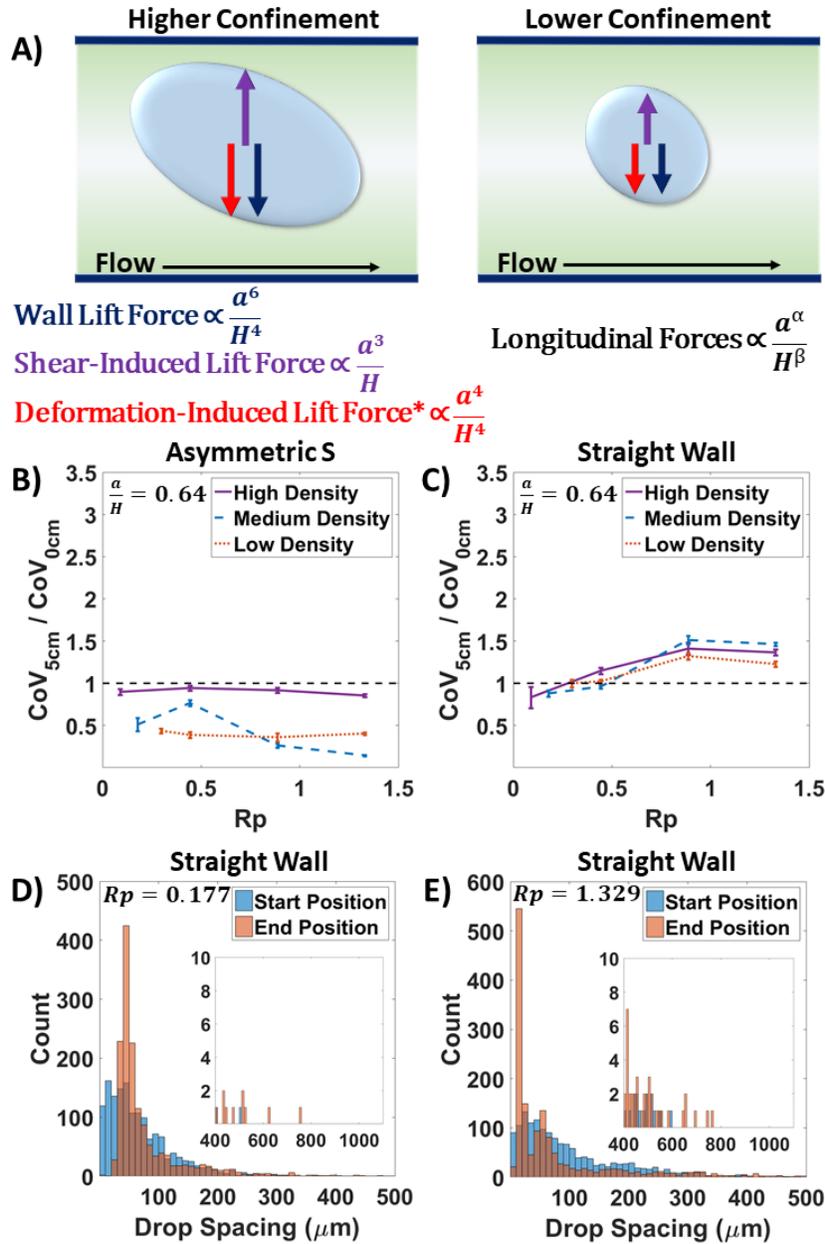

**Figure 7.** Longitudinal ordering at lower confinement ($a/H$ = 0.64, $a$ = **32 μm**). See Supplementary Fig. S8 for detailed $CoV$ plots. Lowest $R_p$ of 0.177 for the medium number density corresponds to a total flow rate of **120 μL·hr$^{-1}$**, and the lowest $R_p$ of 0.296 for the low number density corresponds to **200 μL·hr$^{-1}$**. Example histograms for the straight wall geometry and medium number density are shown. Bin size is **10 μm**. $N$ = **1500** drops for each histogram at each channel position. See Supplementary Fig. S8 for additional histograms. (A) Diagram illustrating that confinement directly affects the magnitude of lift forces as well as the deformation of the drop. Length of arrows only indicate force magnitudes at higher confinement compared to those at lower confinement, not to indicate that lift forces are equal at a given confinement. Wall and shear lift forces scale differently with confinement [2]. Confinement scaling for longitudinal repulsive interactions between drops have not been described analytically or numerically, but the data demonstrates confinement dependent longitudinal ordering. *This analytical solution assumes drops are far from the wall [41,48], thus this previously described confinement scaling is imperfect, particularly with respect to our study. (B) and (C) Summary plots of final spacing improvement, similar to those shown in Fig. 5. $N \approx$ **4500-6000 drops** per data point. (D) and (E) Straight wall geometry at total flow rates of **120** and **900 μL·hr⁻¹**, respectively. Ordering similarly worsens with increasing inertia, as in the 0.8 confinement case. However, the spacing is not as strongly polarized between very small and very large values. In addition, it appears that two peaks form in the highest flow rate case, at spacings < **100 μm**.

     We initially tested drop ordering at a very high confinement ($a/H$ = 0.8), which defines an ordering regime that is both practically relevant to droplet microfluidics as well as being understudied. In this section, we present how confinement alters longitudinal inertial ordering by comparing the same set of experiments using **32 μm** drops ($a/H$ = 0.64), where the diameter change corresponds to nearly halving the drop volume. Despite what is known about the effects of confinement on general lateral lift forces, how longitudinal interactions scale with confinement for drops is undescribed, both analytically and numerically. As illustrated in Fig. 7A, when confinement is decreased, lift forces also weaken. One of these lift forces is the deformation-induced lift force, which scales with confinement [29,30,48] and is weakened for smaller drops as they deform less. Supplementary Fig. S7 shows the lower confinement drops being less deformed at the highest flow rate compared to the higher confinement drops, as characterized with the capillary number. Deformation is directly tied to confinement due to the fact that the shear stress exerted on the drop strengthens with increasing confinement [30]. Simulations have also shown that the local topology of reversing streamlines is altered by drop deformation [49]; the larger the confinement, the greater the deformation, with the reversing streamlines being closer to the more confined and deformed drop [29]. A more deformed drop also has a lower slip velocity and can travel faster down the channel [35], which further influences the dynamics of drop-drop interactions. Coupling to interfacial surfactant effects will be discussed in the following section.

     Fig. 7B and 7C summarize the final spacing improvements using $CoV$ ratios, similar to Fig. 6A and 6B. For the lower confinement drops in the asymmetric serpentine geometry, the high number density case also does not lead to meaningful ordering, with the $CoV$ ratio being close to 1 at all $R_p$ values. The medium number density case shows inertial dependence, with only the highest $R_p$ exhibiting a similarly low $CoV$ ratio (~0.15) compared to the drops at $a/H$ = 0.8 (Fig. 6A). At low number density, there is instead a lack of inertial dependence. The inertial dependence at medium number density and lack thereof at low number density is the opposite of what is observed at $a/H$ = 0.8. However, the precise nature of the inertial dependence is also different, with the $a/H$ = 0.64 drops showing *decreasing $CoV$* with increasing $R_p$ (at medium number density), but the $a/H$ = 0.8 drops showing *increasing $CoV$* with increasing $R_p$ (at low number density). This reversal of sensitivity to inertia is purely due to a change in confinement and highlights the direct influence of this parameter on the longitudinal inertial ordering of drops.

     Drops in the straight wall geometry do not achieve uniform spacing (Fig. 7C) and the *CoV* ratios increase with inertia, rising above 1. But the rise is significantly less compared to that of the drops at $a/H$ = 0.8 (Fig. 6B). Furthermore, the inertial dependence and the $CoV$ ratios themselves are very similar for all three number densities. This relative homogeneity across number densities is, in fact, distinctive compared to the other three confinement/geometry combinations. This distinction is most notably exemplified in Fig. 7D vs 7E (medium number density) and Supplementary Fig. S9A vs S9C (low number density) and S9B vs S9D (high number density). At high $R_p$, there are two notable peaks at the same locations in the histograms: one large peak at the **20 μm** bin and one smaller peak at the **60 μm** bin. The

presence of these two peaks is only seen for the lower confinement drops in the straight wall geometry, and only at higher $R_p$, regardless of number density. Number density primarily manifests inertia-dependent differences in these histograms based on the frequency of large spacing values, which is positively correlated with inertia and negatively correlated with number density. These results suggest that at high $R_p$, there exists 2 number-density-independent, stable fixed points for the lower confinement drops in the straight wall geometry. There is also a possible third fixed point that arises with decreasing number density due to the observation of more frequent large spacing values. The trend of more fixed points appearing when lowering confinement qualitatively agrees with the previously discussed numerical study [50], wherein 3 fixed points—2 stable and 1 unstable—emerge at low confinement. Quantitatively, however, our results differ, as the numerical study predicts 3 fixed points only at $a/H$ < 0.2, and still predicts only a single stable fixed point at $a/H$ = 0.64.

A key influence of confinement on longitudinal inertial ordering is the strength and role of the wall force. We confirm that the drops with both confinements sense the wall force. Supplementary Movie S1 shows that a drop ($a/H \approx 0.6$) moving in the middle of a sawtooth wall geometry senses the wall pattern through induced deformations that result in an undulating drop interface (Supplementary Movie S1). As the wall force is from all directions, it serves to counter "buckling", which is counterproductive to ordering and has been observed in smaller solid particle trains [54] as well as our lower confinement drops (Supplementary Fig. S7). Buckling occurs when the wall force is too weak to prevent the repulsive longitudinal interactions from pushing particles out of a focused single line and into a staggered formation, which is the case at lower confinement. Despite the minimization of buckling for high confinement drops, our observations show that uniform longitudinal ordering could not be achieved in the straight wall geometry (Fig. 6B). In addition, the higher confinement drops at low number density and high $R_p$ (Fig. 6A) in the asymmetric serpentine geometry show worse $CoV$ compared to the lower confinement drops at similar conditions (Fig. 7B). Although Dean flow and deformation complicates buckling for drops in the asymmetric serpentine, overall, our results show that the positive correlation of wall force strength with confinement is not a universal predictor of better longitudinal ordering outcomes when increasing drop confinement.

The question of high confinement, which we have experimentally demonstrated to be a crucial parameter for longitudinal inertial ordering, has not been thoroughly investigated. Numerical treatments of drops moving near walls have not been profuse [44,46,47] and do not adequately incorporate the parameter values relevant to our study and to most droplet microfluidics applications (i.e. high confinement, Marangoni effects, Dean flow, drop-drop interactions). Experimentally, we are not aware of any report investigating the role of confinement on the longitudinal inertial self-assembly of drops. Note that if $a/H \geq 1$, flow becomes constricted and is no longer three-dimensional, which is a flow regime where normal inertial focusing would not apply. Furthermore, such a flow regime can also lead to wall friction effects and unpredictable instabilities [79] that would be highly sensitive to the inevitable fabrication defects inherent to the channel walls.

**D. Surfactant Concentration and Marangoni Forces Change Ordering Dynamics**

Marangoni forces stem from surface tension gradients that can arise due to the uneven distribution of surfactants at the interface between two fluids. Given that microfluidic drops are made using surfactant and subject to shear flow, this could induce a rearrangement of surfactant molecules leading to surface tension gradients. These gradients lead to interfacial flows, for which the approximate speed $v$ can be estimated by:

$$v \approx \frac{\Delta\sigma}{\mu}, \tag{12}$$

Given that $\mu$ is of order **1 mPa·s** and $\sigma$ is of order **10 mN·m⁻¹** here (see Appendix), a ~10-50% differential in $\sigma$, which simulation has shown can be within that range [36], can lead to Marangoni flow speeds of order **1-5 m·s⁻¹**. These flows can lead to viscous interactions with the surrounding fluid (Fig. 8B), which have been described as a potential "swimming" action in drops [41]. By dimensional analysis, the magnitude of these viscous forces $F_f$ can be estimated using $\mu$, $v$, and the length scale of the drop $a$:

$$F_f \sim \mu v a, \tag{13}$$

For a drop diameter of **40 μm** and the range of $v$ estimated above, $F_f$ is of order **40-200 pN**, which is comparable to the lift force given in Eq. 11 at the lowest $R_p$ (see Supplementary Material), making these forces a non-negligible factor. The influence of such Marangoni effects on the lateral migration of drops has been documented experimentally [41,42], but they could conceivably affect the longitudinal many-body dynamics of drops as well, which has not been demonstrated. To test this hypothesis, drops of both sizes were made using 0.5% surfactant ($\sigma =$ **14.58(3) mN·m⁻¹**) to ensure an appreciable effect without risking significant drop coalescence during reinjection, whereas all previous experiments used the standard 2% common in droplet microfluidics ($\sigma =$ **9.44(1) mN·m⁻¹**). Comparison was done using only the asymmetric serpentine geometry as the straight wall geometry previously demonstrated an inability to longitudinally order drops. Fig. 8 plots the position-by-position ratio of the $CoV$'s measured for the 0.5% drops to those previously measured for the 2% drops. Thus, a $CoV$ ratio > 1 would indicate poorer ordering due to less surfactant. Only the medium and low number densities were examined since the high number density condition fails to achieve meaningful ordering.

      Our observations suggest that the effects of Marangoni forces on longitudinal inertial ordering are less pronounced at higher confinement than at lower confinement. Fig. 8C vs 8E (medium number density) and Fig. 8D vs 8F (low number density) show that there are greater overall deviations from a $CoV$ ratio of 1 at lower confinement. In particular, decreasing surfactant concentration tends to worsen performance (higher $CoV$ ratio) at lower confinement. These results indicate that at least for the surfactant concentrations tested here—which are in a range typical of droplet microfluidics—confinement is the more dominant factor.

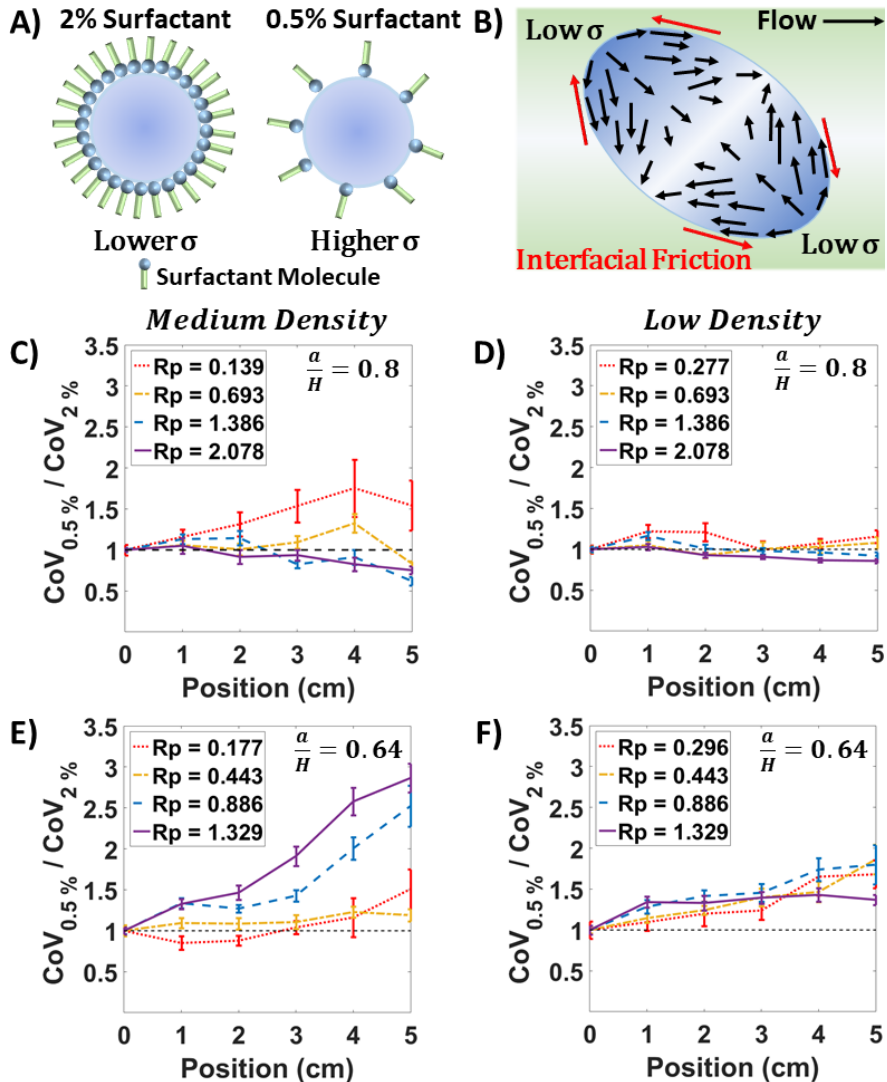

**Figure 8.** Results demonstrating the impact of surfactant concentration and the Marangoni effect on longitudinal inertial drop ordering. Channel geometry is asymmetric serpentine for all cases. $N \approx$ **4500-6000** drops per data point. The data points are the ratio of the $CoV$ for 0.5% surfactant drops (higher surface tension, $\sigma$) to the $CoV$ of 2% surfactant drops at each channel position. A $CoV$ ratio above 1 indicates poorer longitudinal ordering performance at 0.5% surfactant. See Supplementary Fig. S10 and S11 for individual plots. (A) Sketch indicating that lower surfactant concentration leads to less available coverage of the drop. Amount of surfactant molecules depicted is exaggerated and not an accurate representation of surface saturation percentage. (B) Sketch depicting that deformed drops in flow tend to have surfactant more localized at the two tips [36], creating low surface tension poles that result in surface tension gradients (black arrows). These gradients can result in interfacial flow away from the drop tips, potentially generating viscous forces against this flow, shown with red arrows [41]. (C) and (D) contrasts the effect of number density on the comparison between 0.5% and 2% surfactant drops at $a/H$ = 0.8. (E) and (F) contrasts the two surfactant concentrations at the lower confinement of $a/H$ = 0.64. Here, stronger inertial dependence emerges and $CoV$ ratios tend to be higher.

For the inertial dependence aspect, the data shows that it is influenced differently by confinement based on the number density. At low number density, decreasing confinement causes the $CoV$ ratio to trend noticeably above 1 at all $R_p$, but the increase is similar and the curves overlap (Fig. 8F). At medium number density, however, changing confinement leads to drastically divergent outcomes at the higher $R_p$ values. At $a/H$ = 0.8 (Fig. 8C), the $CoV$ ratio trends below 1 at $R_p$ > 0.7, implying that any negative effects on longitudinal ordering due to a lower surfactant concentration are eliminated at higher inertia under high confinement. This stands in contrast to the $a/H$ = 0.64 case (Fig. 8E), where the $CoV$ ratio rises to between 2 and 3 at $R_p$ > 0.7, indicating a confinement-dependent reversal of the influence on longitudinal ordering due to a lower surfactant concentration. As a lower concentration equals a higher average surface tension, the medium number density results seem to suggest that making drops less deformable in the face of higher inertia—which increases deformation—is conducive to longitudinal ordering at higher confinement but detrimental to it at lower confinement. This observation may imply that an optimal degree of deformation exists at each confinement, and at higher $R_p$, it lies above the degree experienced by 2% surfactant drops at $a/H$ = 0.64 confinement but below that experienced by 2% drops at $a/H$ = 0.8.

The dependence of the deformation parameter on surfactant concentration is further complicated by effects that extend beyond the simple modulation of average surface tension. The degree of drop deformation can change the sign of the normal interfacial stresses [80] and couple to the spatial distribution of surfactants, which gives rise to the tangential Marangoni stresses. A study [36] has shown that surfactant tends to concentrate at the tips of drops under shear flow, which favors greater deformation. The deformation-confinement coupling also feeds this surfactant distribution aspect as confinement affects the inclination angle of the drop with respect to the axis of flow [28], and this determines how pointed the drop tips are [29]—which is where the surfactant is concentrated. Curiously, a simulation study shows that the resultant Marangoni stresses, which form surface tension gradients pointing away from the drop tips and toward the drop belly (Fig. 8B), could have a negligible impact on the deformation [36]. However, this study simulated perfectly centered drops with viscosity and mass density ratios of 1. Nonetheless, it is the spatial distribution of surfactants and the accompanying Marangoni effects that is significant, as despite the fact that $\sigma$ is within a factor of two for 0.5% and 2% surfactant drops, $\Delta\sigma$ and therefore $F_f$ can be dynamically and appreciably different.

We have presented the first experimental evidence of surfactant and Marangoni effects influencing the longitudinal inertial self-assembly of microfluidic droplet arrays. Considering the core usage of surfactants in droplet microfluidics, knowledge of this parameter's effects on self-ordering is pivotal for integrating inertial principles into droplet microfluidics.

## IV. Conclusion

In this study, we have experimentally shown that even when using conventional channel geometries, the unique properties of droplets can yield unexpected results in inertial ordering compared to solid particles, such as the failure of the traditional straight wall geometry to achieve uniform spacing—counter to established paradigm. On the other hand, the asymmetric serpentine geometry was able to restore regular drop spacing over length scales at least 3 orders of magnitude greater than the drop diameter, especially at low inertia. We *independently* varied the parameters of channel geometry, number density, confinement, inertia, and surfactant concentration to investigate how they can affect the longitudinal inertial ordering of drops. However, many questions have been raised through our work regarding the complex nature of these fluid interactions, which cannot be adequately explained by existing literature. To gain a better understanding of the underlying physics, it seems that new computational efforts will be required to grapple with these multiphase fluid systems. A recent review [81] has indicated that key areas are still missing in the computational research surrounding the inertial ordering of deformable particles, particularly under high confinement, Dean flow, and the influence of interfacial effects. But we hope this work inspires and provides the experimental impetus for future theoretical and computational endeavors. Experimentally, we will continue to study this system to evaluate the stability these droplet arrays under perturbations and to expand this work toward non-Newtonian fluids. In the meantime, our results can still serve to inform on the utility of incorporating inertial ordering principles for direct application in droplet microfluidics. Our identification of conditions yielding stable, long-range, and passive ordering of drops achievable at the range of flow rates compatible with droplet microfluidics enables the simple integration of incubation and deterministic interrogation modules. These capabilities should prove useful for streamlining multiplexed and multistep droplet microfluidic assays.

## Acknowledgments

This work was funded by the UIUC Startup Grant, with fabrication and physical characterization carried out in part in the Materials Research Laboratory Central Research Facilities, University of Illinois.

## Appendix

### A. Materials

All devices (drop makers and inertial ordering chips) are made with a single layer of polydimethylsiloxane (PDMS) mixed at a ratio of **1:10** (curing agent to base) and plasma bonded to a glass slide. These processes employ the standard methods of soft and photolithography, with master molds made using SU-8 2050 photoresist. Channels are made hydrophobic through Aquapel™ treatment. In all experiments, flow is driven by syringe pumps (Poiseuille flow). Aqueous drops serve as the dispersed phase and constitute water with fluorescein isothiocyanate (FITC) dye while the continuous phase is hydrofluoroether oil (Novec™ HFE 7500), which is the denser ($\rho$ = **1.61 g·cm$^{-3}$**) and more viscous fluid ($\mu$ = **1.24 mPa·s**). For drop making, the continuous phase (HFE) also contains fluorosurfactant (RAN Biotechnologies™ 004B) dissolved at either 2% or 0.5% w/v.

# Supplementary Material

Wenyang Jing and Hee-Sun Han

## Buoyancy vs Net Lift Force Calculation

According to Stan et al. [1], the net empirical lift force on a drop, directed toward the channel center, is given by:

(14)
$$F_{L,empirical} = C_L \mu U r \left(\frac{r}{H}\right)^3 \left(\frac{d}{H}\right),$$

Here, $C_L$ is empirical lift coefficient, which the authors recommend to be 500 for general order of magnitude estimates, $\mu$ is the dynamic viscosity of the carrier fluid, which $= 1.24 \text{ mPa} \cdot \text{s}$ for HFE oil, and $U$ is the average velocity of the fluid in the channel, which for the purposes of this calculation will correspond to the lowest flow rate used: $60 \, uL \cdot hr^{-1}$ and $U \approx 0.667 \text{ cm} \cdot s^{-1}$. $H$ is the hydraulic diameter (**50 μm**), $r$ is the drop radius, and $d$ is the distance from the center of the channel. The empirical lift force will be computed for **40 μm** and **32 μm** drops in the scenario where they are positioned just so they are midway between the channel center and the upper channel wall. This scenario is to confirm if the buoyant force is sufficient to make the drops rise to this midpoint. Therefore, $d/H$ will be 0.05 and 0.09 for **40 μm** and **32 μm** drops, respectively.

The net upward force resulting from the difference in the buoyant force and the weight of a water drop is given by:

(15)
$$F_{B,net} = \left(\frac{4\pi}{3}\right) r^3 g (\rho_{out} - \rho_{in}),$$

Here, $g = 9.81 \, m \cdot s^{-2}$ is the gravitational acceleration, $\rho_{out} = 1614 \, kg \cdot m^{-3}$ is mass density of the carrier fluid, HFE, and $\rho_{in} = 1000 \, kg \cdot m^{-3}$ is mass density of the water drop.

**40 μm** drops:

$$F_{L,empirical} = 500 * 0.00124 * 0.00667 * 20 * 10^{-6} \left(\frac{20}{50}\right)^3 \left(\frac{2.5}{50}\right) = \mathbf{265 \, pN}$$

$$F_{B,net} = \left(\frac{4\pi}{3}\right)(20 * 10^{-6})^3 * 9.81 * (1614 - 1000) = \mathbf{202 \, pN}$$

**32 μm** drops:

$$F_{L,empirical} = 500 * 0.00124 * 0.00667 * 16 * 10^{-6} \left(\frac{16}{50}\right)^3 \left(\frac{4.5}{50}\right) = \mathbf{195 \, pN}$$

$$F_{B,net} = \left(\frac{4\pi}{3}\right)(16 * 10^{-6})^3 * 9.81 * (1614 - 1000) = \mathbf{103 \, pN}$$

The empirical lift force assumes a capillary number and Reynolds number < 0.01, but only the capillary number condition is satisfied. Nonetheless, this estimate is still useful to show that the net upward force from buoyancy should not be sufficient to push the drops up against the upper wall of the channel as the lift force will only get stronger with higher inertia while the buoyant force will remain constant. Lateral equilibrium positions affected by buoyancy, but is also a natural part of droplet microfluidic systems, where density matching is impractical. How buoyancy affects the lateral dynamics in the asymmetric serpentine geometry is more complicated due to the influence of Dean flow, but given its success over the straight wall geometry, this is a lesser practical concern.

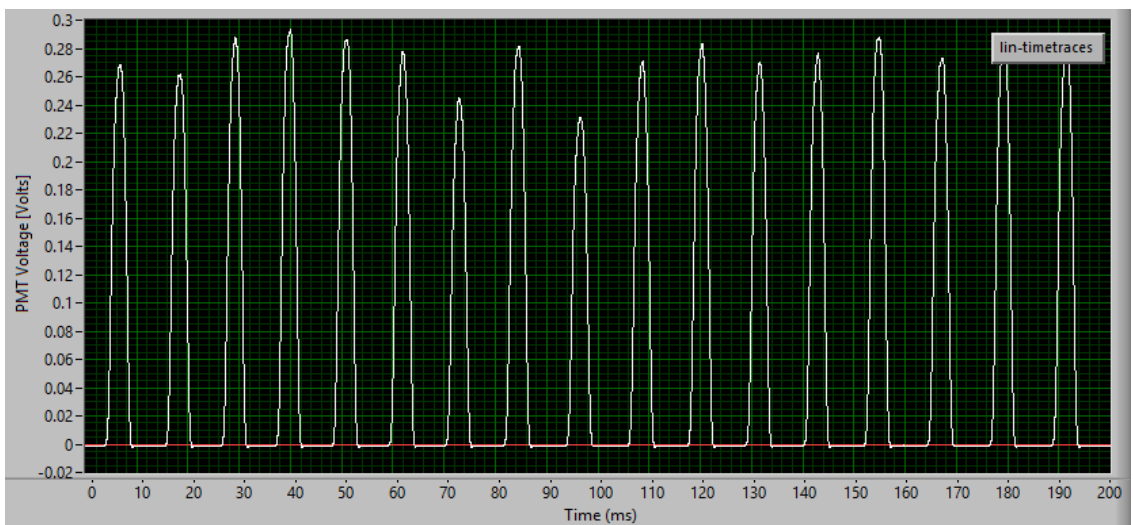

**Fig. S1.** Example PMT trace of fluorescein drops, as seen in LabVIEW. Peak width and spacing between peaks are recorded as time values (**ms**).

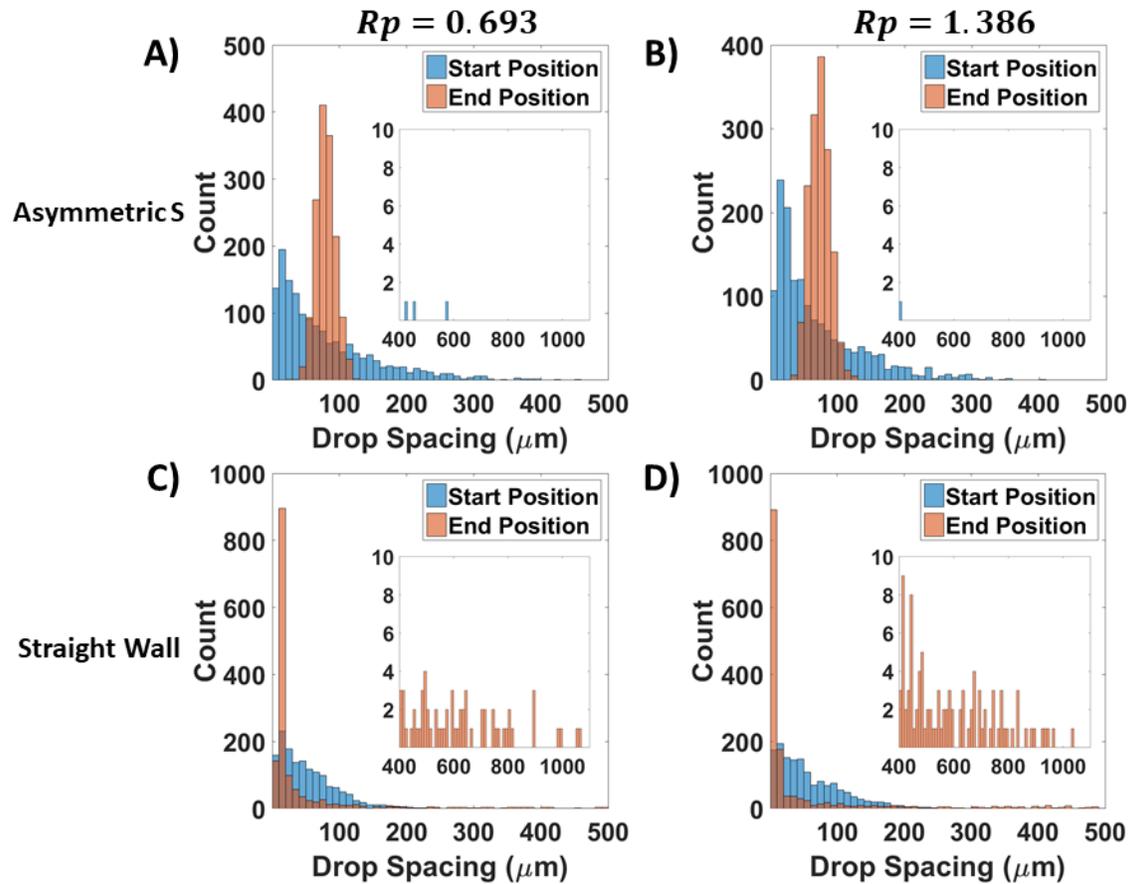

**Fig. S2.** Example histograms of spacing distribution at the medium number density, confinement of 0.8 (**40 μm** drops), and 2% surfactant. Bin size is **10 μm**. Start and End Positions refer to **0** and **5 cm** locations in the channel. $N$ = **1500** drops in all graphs, for each respective histogram. (A) and (B) Asymmetric serpentine geometry and total flow rates of **300** and **600 μL·hr$^{-1}$**, respectively. (C) and (D) Straight wall geometry and total flow rates of **300** and **600 μL·hr$^{-1}$**, respectively. Note that in this geometry, not only are larger spacings more prevalent in the end position histogram, but the number of short spacings also increases with inertia.

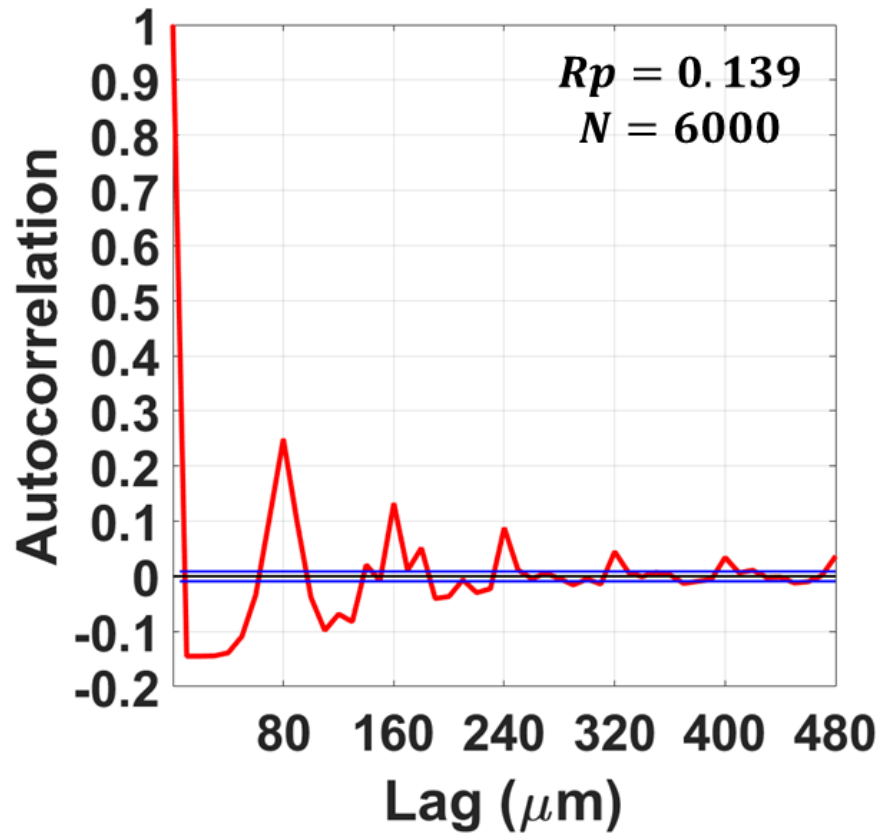

**Fig. S3.** The autocorrelation function for the combination of 4 separate data files each containing $N = $ **1500** drops for the asymmetric serpentine geometry, $a/H = 0.8$, and 2% surfactant. As these 4 data files were taken at different points in time, covering approximately 20 minutes of flow at this condition, the periodic nature still evident here demonstrates stability in the system with time. Blue lines represent 95% confidence intervals.

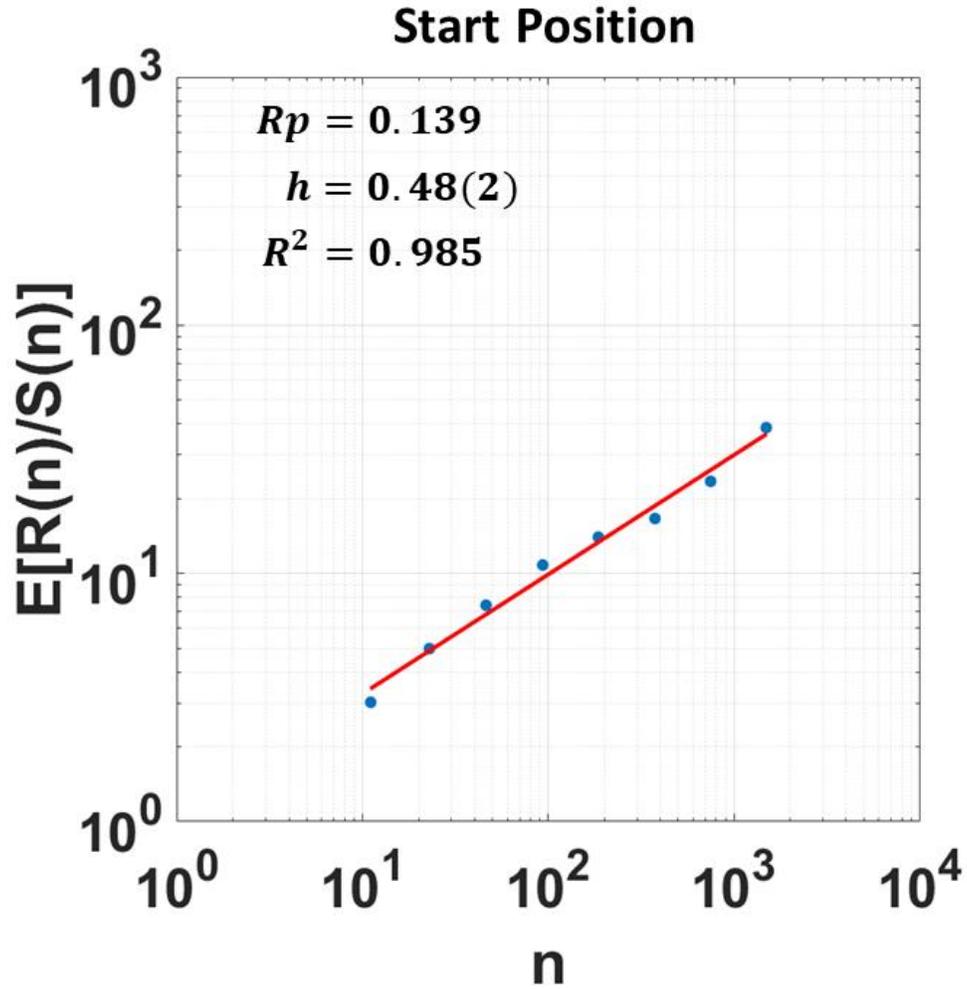

**Fig. S4.** Example rescaled range analysis for drops at the start position in the channel following the spacing disruptor, which is the same initial condition for both channel geometries. Drop distribution is indeed randomized and uncorrelated, as confirmed by the Hurst exponent being 0.48(2), where 0.02 is the standard error of the slope on the log-log plot.

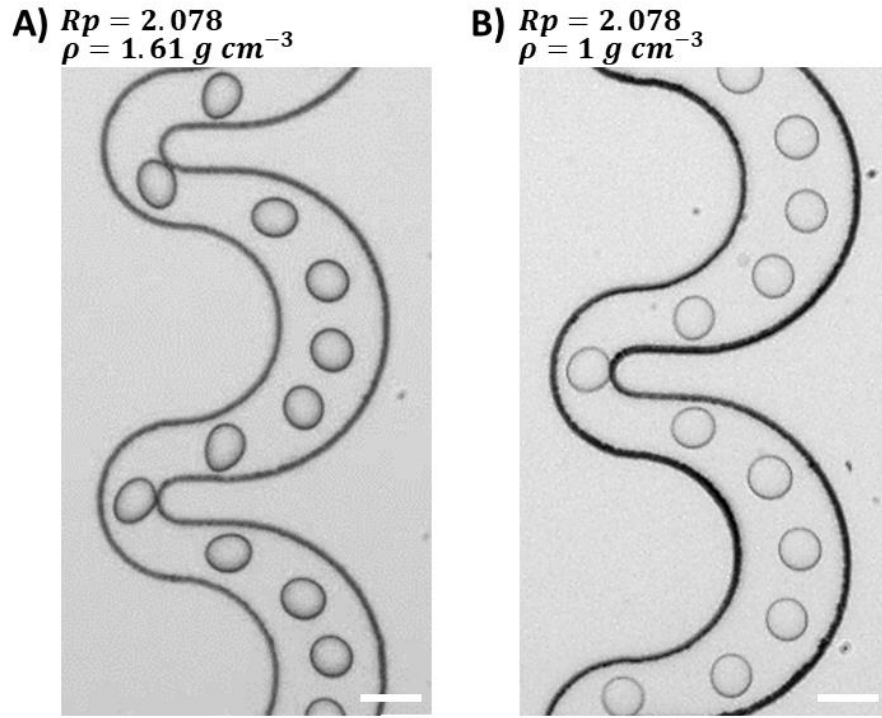

**Fig. S5.** Images of drops flowing the asymmetric serpentine geometry at a total flow rate of **900 µL·hr⁻¹**, $a/H$ = 0.8, and 2% surfactant. (A) Drops that have their mass density matched to HFE oil ($\rho$ = **1.61 g·cm⁻³**) by dissolving cesium chloride salt. Note that the drops are more deformed. (B) The aqueous drops used in this study without mass density adjustment. This indicates that viscosity in the drops has changed, and unfortunately shows that the drops cannot be made neutrally buoyant without affecting other properties, such as the viscosity ratio. Scale bars are **50 µm**.

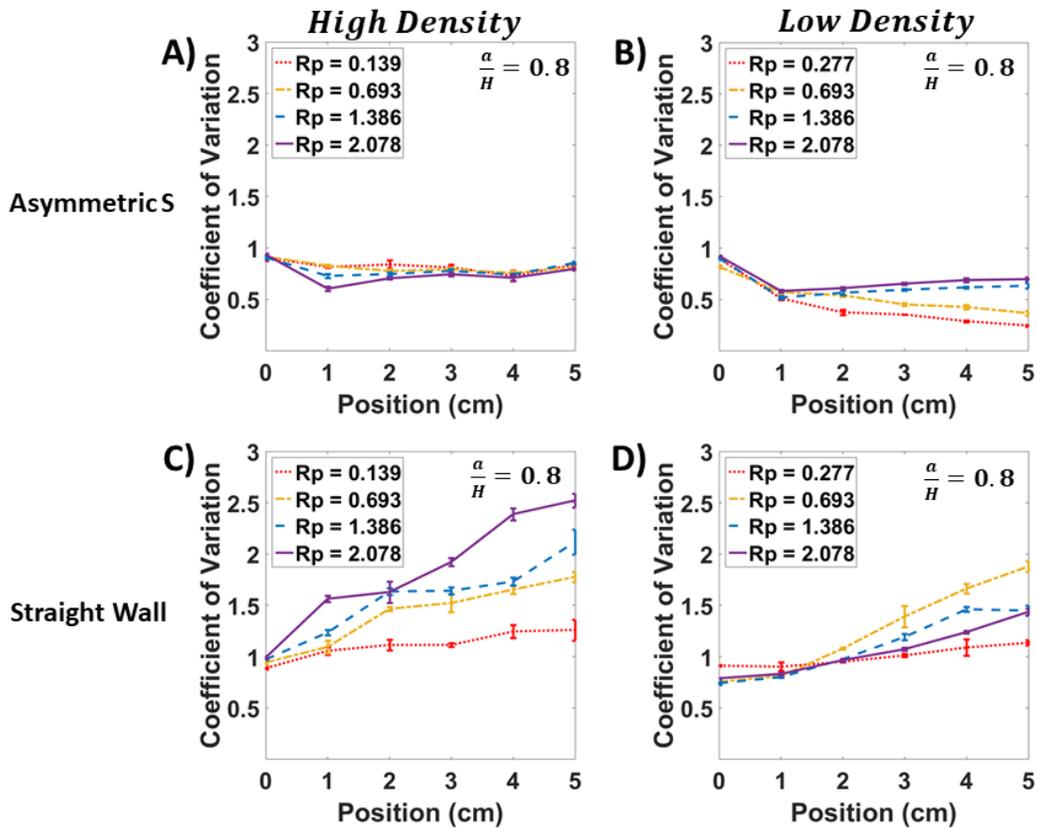

**Fig. S6.** Individual *CoV* plots for the higher confinement case ($a/H$ = 0.8), complementary to plots shown in Fig. 2 and 6. $N \approx$ **4500-6000** drops per data point. High number density is ~**7400 drops·µL$^{-1}$**, and low number density is ~**1900 drops·µL$^{-1}$**. (A) Asymmetric serpentine geometry, no meaningful longitudinal ordering achieved despite the channel geometry, likely too much steric interference at high number density. (B) Asymmetric serpentine geometry, divergent inertial dependence at low number density, in contrast to Fig. 2A. (C) Straight wall geometry, at high number density, spacing distribution reflecting drops becoming more clustered with short drop spacings within clusters and large gaps between clusters. (D) Straight wall geometry, the lower number density weakens longitudinal interactions and so the coefficient of variation does not increase as much as the medium and high number density cases. There is also less inertial dependence.

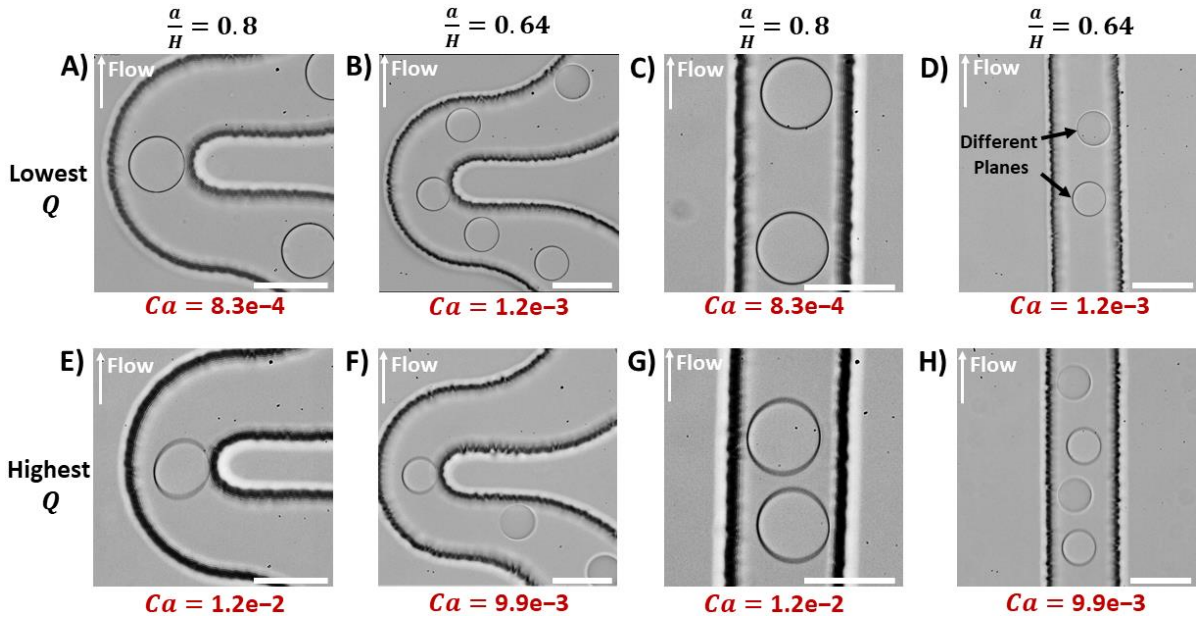

**Fig. S7.** Images of drops at lowest and highest total volumetric flow rates ($Q$) used, showing the extent of drop deformation and the lateral organization of the drops at both confinements. The highest $Q$ for both confinements is **900 μL·hr⁻¹**, the lowest $Q$ is **60 μL·hr⁻¹** and **120 μL·hr⁻¹** for the confinements of 0.8 and 0.64, respectively. Capillary number $Ca$ is defined as $\mu U a/(\sigma y)$ [2], where $\mu$ is the dynamic viscosity of the continuous phase, $U$ is the average fluid velocity, $a$ is the drop diameter, $\sigma$ is the surface tension, and $y$ is the channel height. (A) and (E) Drops always hug the inner radius as it rounds the smallest curve in the asymmetric serpentine, see Movie S2. The drop is visibly non-spherical at the highest $Q$. (B) and (F) Smaller drops also prefer the inner radius, but are not as deformed at the highest $Q$. They can also be seen to be in different focal planes. (C) and (G) At the lowest $Q$, the drops at this confinement prefer a more centered lateral alignment. But at the highest $Q$, drops tend to be closely clustered and off-centered in their lateral alignment, which also deforms them more due to a more uneven shear as a result of their off-center position. (D) and (H) The smaller drops do not experience a specific inertia-dependent lateral alignment, unlike at higher confinement. Scale bars are **50 μm** for all images.

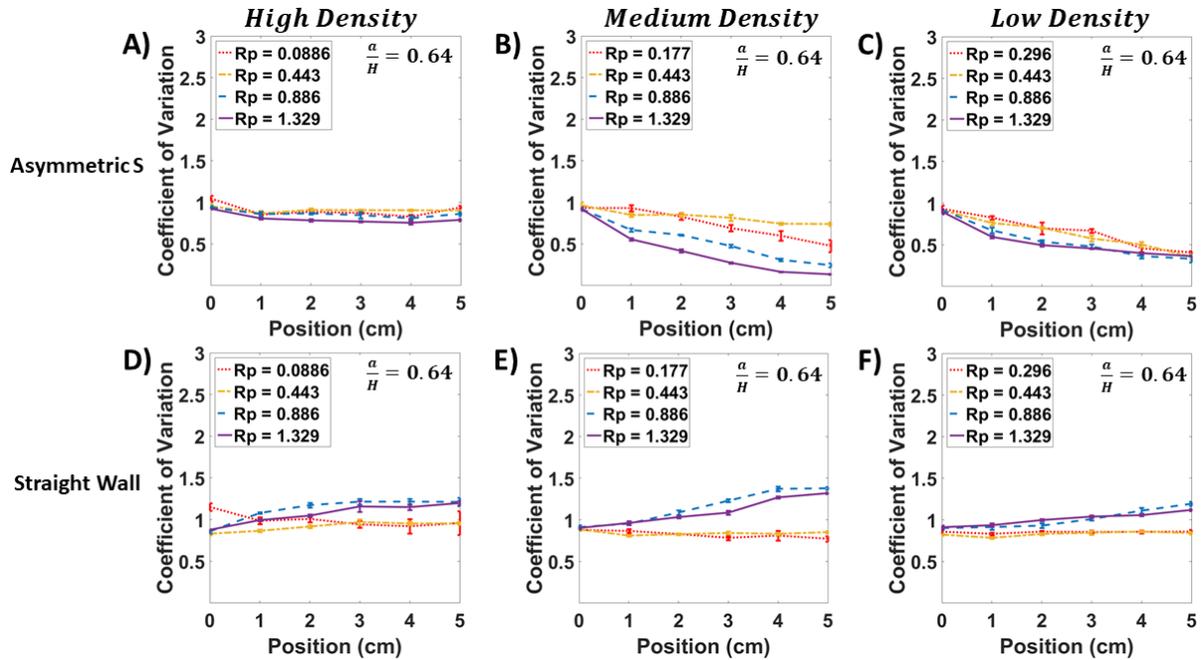

**Fig. S8.** This figure shows the explicit spacing characterization data as a function of downstream channel position for the lower confinement case. $R_p$ values of 0.443, 0.886, and 1.329 correspond to total flow rates of **300**, **600**, and **900 µL·hr$^{-1}$**, respectively. The $R_p$ of 0.177 for the medium number density case corresponds to **120 µL·hr$^{-1}$**, and the $R_p$ of 0.296 for the low number density case corresponds to **200 µL·hr$^{-1}$**. This was necessary as the aqueous flow rate could not be reliably lowered below **10 µL·hr$^{-1}$**, so the oil flow rate had to be increased to decrease number density. $N \approx$ **4500**-**6000** drops per data point, with error bars being the standard deviation from averaging separate trials at $N \approx$ **1500** drops each. The summary results are presented in Fig. 7, where the ratio of the CoV at the **5 cm** position to **0 cm** position for each is plotted. (A)-(C) Asymmetric serpentine. Medium number density achieves the best possible result for $R_p$ = 1.329, and shows inertial dependence. (D)-(F) Straight wall geometry. No meaningful ordering is achieved. Also exhibits the greatest inertial dependence—split between the upper and lower $R_p$ pairs—at the medium number density. Increases in $CoV$ do not rise as much compared to $a/H$ = 0.8 in the straight wall geometry.

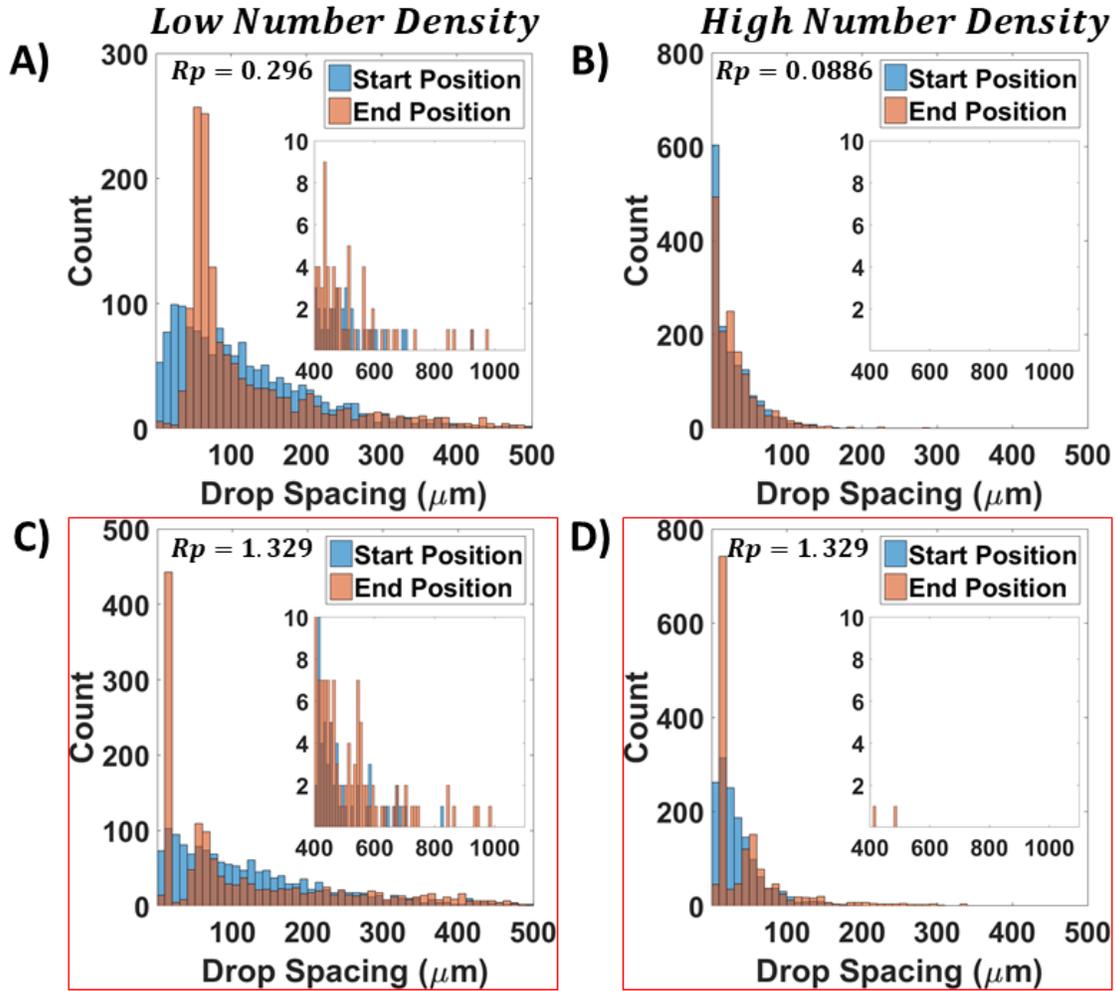

**Fig. S9.** Example histograms for the **32 μm** drops ($a/H$ = 0.64) in the straight wall geometry at the lowest and highest flow rates tested and for the low and high number densities. $R_p$ values of 0.0886, 0.296, and 1.329 correspond to total flow rates of **60**, **200**, and **900 μL·hr⁻¹**, respectively. $N$ = **1500** drops for each histogram and bin size is **10 μm**. (A) and (C) Low number density cases, note the presence of a single peak at lower $R_p$ but the presence of two peaks at $R_p$ = 1.329 in exactly the same positions as shown in Fig. 7E, which is the medium number density. (B) and (D) High number density cases, where at low $R_p$, the histograms at the start and end positions are very similar, but at high $R_p$, there are again both peaks in the same position as was shown in Fig. 7E and here in C.

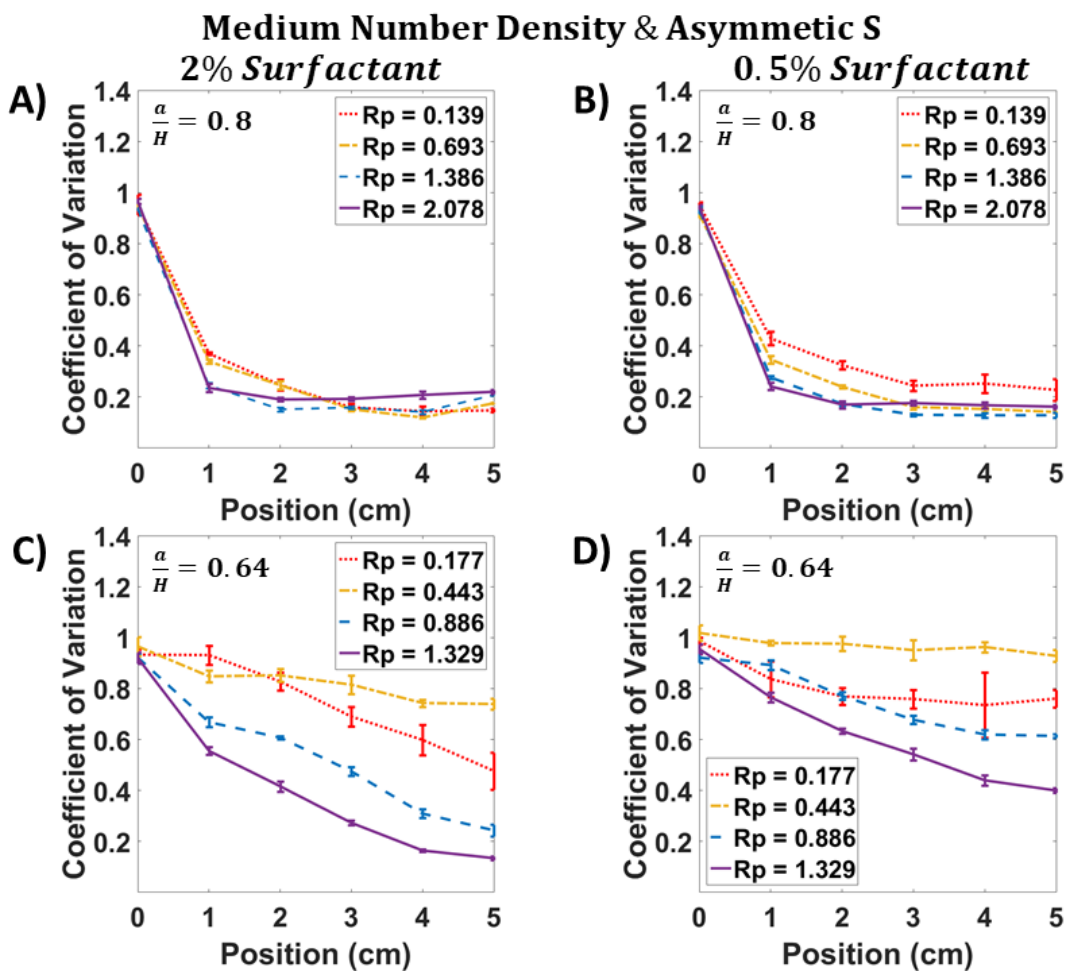

**Fig. S10.** Asymmetric serpentine geometry and number density is ~**3700 drops·µL$^{-1}$** for all graphs. $N \approx$ **4500-6000** drops per data point. (A) and (C) rescaled versions of Fig. 2A and S8B, reproduced here for comparison purposes. (B) and (D) are same size drops made using 0.5% surfactant. Influence of surfactant effects and Marangoni forces is confinement dependent, with greater $R_p$ sensitivity at lower confinement (C vs D). In particular, better longitudinal ordering is positively correlated with increasing $R_p$ and increasing surfactant concentration at lower confinement but more weakly at higher confinement.

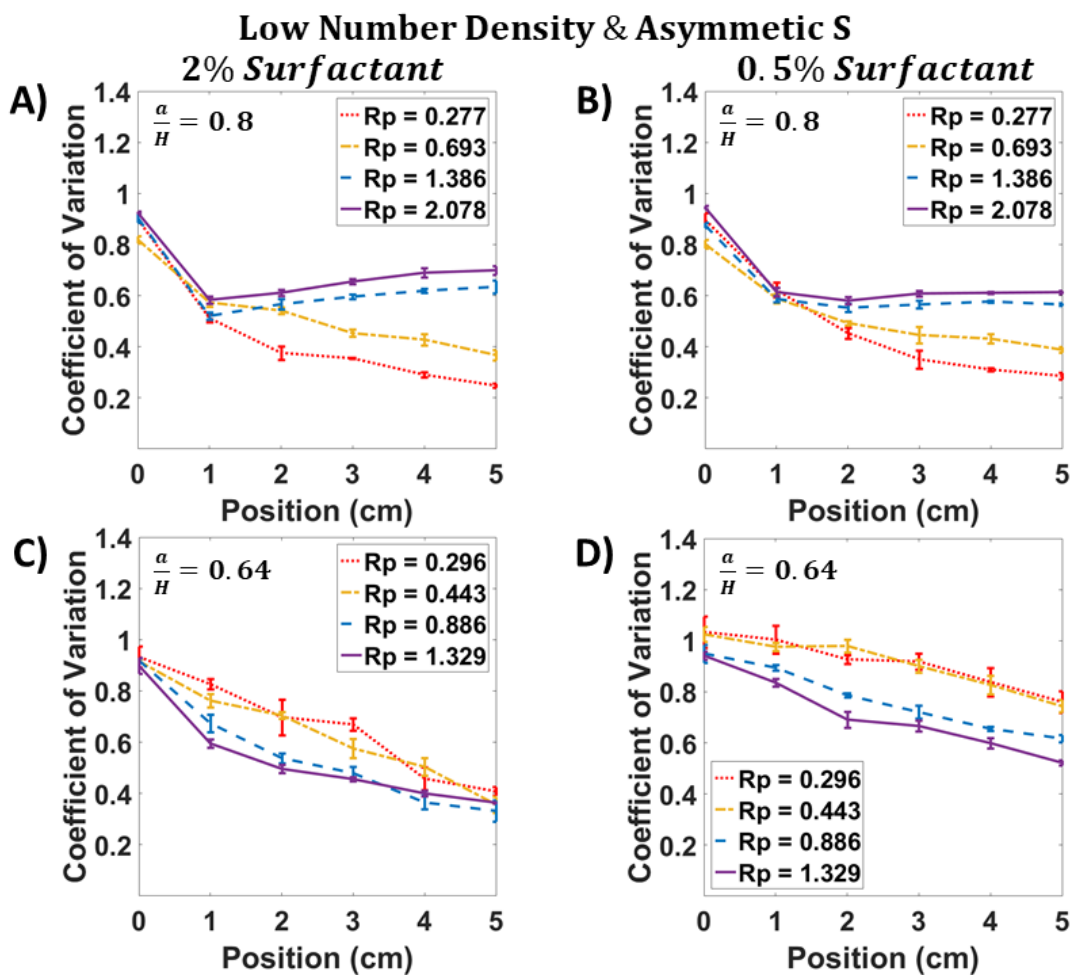

**Fig. S11.** Asymmetric serpentine geometry and number density is ~**1900 drops·µL⁻¹** for all graphs. $N \approx$ **4500-6000** drops per data point. (A) and (C) are rescaled from Fig. S6C and S8C. (B) and (D) are same sized drops made using 0.5% surfactant. At lower surfactant concentration, the divergent behavior at low vs high $R_p$ in A is weakened compared to B. For lower confinement (C vs D), $CoV$ improvement with $R_p$ is lessened across the board when Marangoni forces are weaker, with emergent disparity between the lower and higher $R_p$ values.

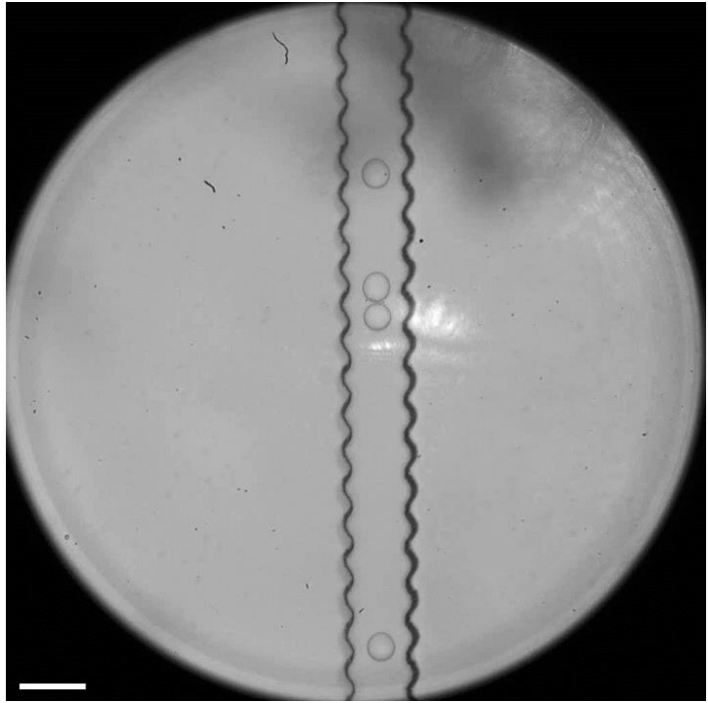

**Movie S1.** This video shows strength of the wall influence, keenly felt by drops at a confinement of ~0.6. The drops' edges undulate in response to the periodic sawtooth wall. Scale bar is **60 µm**.

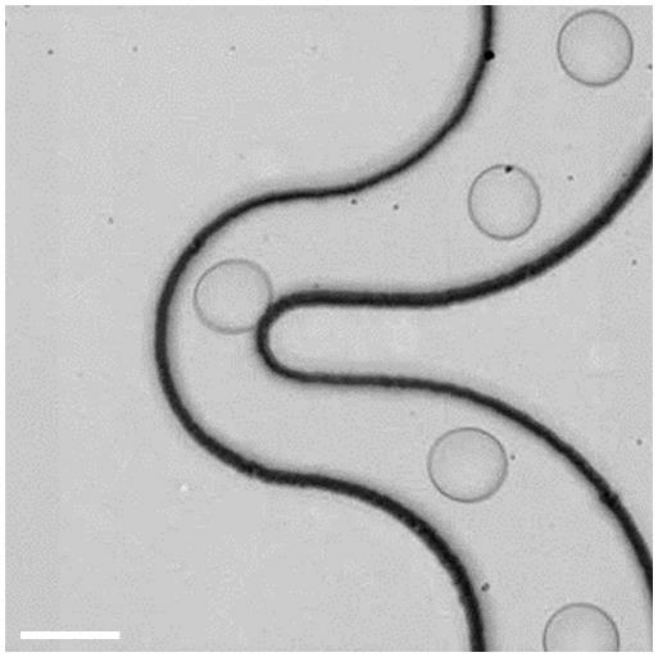

**Movie S2.** This shows drops at $a/H$ = 0.8 and $R_p$ = 2.078. Note that the drops prefer to hug the inner radius on the smallest of the two curves in the asymmetric serpentine. This is true at all the flow rates tested, and both confinements. Scale bar is **50 µm**.